\pdfoutput=1

\RequirePackage{fix-cm}
\RequirePackage{rotating}

\documentclass[smallextended]{svjour3}
\smartqed 
\usepackage{cite}
\usepackage{mathptmx}
\usepackage{rotate}
\usepackage{amsmath,amssymb,amsfonts}
\usepackage{algorithmic}
\usepackage{graphicx}
\usepackage{subcaption}
\usepackage{textcomp}
\usepackage{xcolor}
\usepackage{listings}
\usepackage{multirow}
\usepackage{xcolor,colortbl}
\usepackage[flushleft]{threeparttable}
\usepackage{amsmath}
\usepackage{relsize}
\usepackage{pifont}
\usepackage{wrapfig}
\usepackage{url}
\usepackage[sort&compress,numbers]{natbib}
\usepackage[bottom]{footmisc}

\usepackage{wasysym}

\usepackage{pgfplots}

\usepackage{changepage} 

\pgfplotsset{compat=1.10}
\usepgfplotslibrary{fillbetween}

\def\BibTeX{{\rm B\kern-.05em{\sc i\kern-.025em b}\kern-.08em
    T\kern-.1667em\lower.7ex\hbox{E}\kern-.125emX}}
    
\definecolor{MyDarkBlue}{rgb}{0,0.08,0.45} 
\usepackage{pifont}
\usepackage{listings}
\usepackage{tcolorbox}
\newtcolorbox{blockquote}{colback=red!5!white,boxrule=0.4pt,colframe=red!50!black,fonttitle=\bfseries,top=2pt,bottom=2pt}

\newcommand{\bi}{\begin{itemize}}
\newcommand{\ei}{\end{itemize}}

\usepackage{tabularx}
\usepackage{multirow}
\usepackage{color,colortbl}
\definecolor{lightgray}{gray}{0.8}
\usepackage{adjustbox}

\usepackage{amssymb}
\usepackage{pifont}
\newcommand{\cmark}{\ding{51}}%

\newcommand{\fig}[1]{Figure~\ref{fig:#1}}

\newcommand{\tbl}[1]{Table~\ref{tbl:#1}}

\begin{document}

\title{
Predicting Health Indicators for Open Source Projects \\(using Hyperparameter Optimization)}

 
 
\author{Tianpei Xia   \and
        Wei Fu        \and
        Rui Shu        \and
        Rishabh Agrawal       \and
        Tim Menzies
}

\institute{Tianpei Xia, Wei Fu, Rui Shu, Rishabh Agrawal, Tim Menzies\at
              Department of Computer Science, North Carolina State University, Raleigh, NC, USA \\
              Email: txia4@ncsu.edu, fuwei.ee@gmail.com, rshu@ncsu.edu, agrawa3@ncsu.edu, timm@ieee.org 
}

\date{Received: date / Accepted: date}


\maketitle

\begin{abstract}
Software developed on public platform is a source of data that can be used to make predictions about those projects. 
While the individual developing activity may be random and hard to predict, the developing behavior on project level can be predicted with good accuracy  when large groups of developers work together on software projects.

To demonstrate this, we use 64,181 months of data from 1,159 GitHub projects to make various predictions about the recent status of those projects (as of April 2020). We find that traditional estimation algorithms make many mistakes. Algorithms like $k$-nearest neighbors (KNN), support vector regression (SVR), random forest (RFT), linear regression (LNR), and regression trees (CART) have high error rates.
But that error rate can be  greatly reduced using hyperparameter optimization.

To the best of our knowledge, this is the largest study yet conducted, using recent data for predicting multiple health indicators of open-source projects. 



\end{abstract}

\newpage

\section{Introduction}
\label{sect:intro}
In 2020, open-source projects dominate the software development environment \cite{Paasivaara18,santos2016investigating,Hohl18,Parnin17}.
Over 80\% of the software in any technology product or service are now open-source~\cite{zemlin2017}. 
With so many projects now being open-source, a natural next question is ``which of these projects are any good?'' or ``which should I avoid?''.
 In other words, we now need to assess
 the   {\em health condition} of open-source projects before using them.
Specifically, software engineering
managers need {\em project health indicators} that  assess the health of a project
at some future point in time.

 To assess project health, we look at project activity.
  Han et al. note
 that popular open-source projects tend to be more active~\cite{han2019characterization}. Also, many other researchers agree that healthy open-source projects need to be ``vigorous''~\cite{wahyudin2007monitoring,jansen2014measuring,manikas2013reviewing,link2018assessing,wynn2007assessing,crowston2006assessing}.  
 In this paper, we use   64,181 months of data from  GitHub to make predictions for the   April 2020 activity within 1,159 GitHub projects. Specifically: 
 \begin{enumerate}
 \item The number of contributors who will work on the project; 
 \item The number of commits that project will receive;
 \item The number of opened pull-requests in that project;
 \item The number of closed pull-requests in that project;
 \item The number of opened issues in that project;
 \item The number of closed issues in that project;
 \item Project popularity trends (number of GitHub ``stars'').
 \end{enumerate}

We explore these aspects since,
after several months of weekly
teleconferences with Linux developers, these were issues that many of them were concerned with.
That said,
we make no claim
that this is a complete set of project health
indicators-- after all, currently there is no unique and consolidated definition of project health~\cite{link2018assessing}. However, what we can show is that we have prediction methods that
work well, for thousands of projects,
for all  the indicators shown above. Hence we predict  (but cannot absolutely prove) that
when new indicators arose, our methods will be useful.   

A second result we offer is that, based
on user studies with domain experts from those projects, we can assert that most of the above
indicators are of active interest to the practitioner community. To be sure, some are seen to be of lesser importance (e.g. 52\% of our experts didn't care about ``stars'')
but others are deemed to be very important (e.g.
``close issues'' was described as ``very important''
or ``somewhat important'' by 93\% of our experts).
For the practical purposes, the domain experts consider
the abnormal values of these indicators 
show issues that deserve a manager's attention (and we note that, in the literature, other researchers argue that such abnormal values offer useful guidance for
directing manager intervention~\cite{nagy2010bayesian}).

A third result is that
we demonstrate that it is possible to accurately predict  health indicators for 1, 3, 6, and 12 months into the future. To the best of our knowledge,
no such prediction has been shown possible for
thousands of projects.
 
As to ``how'' we achieve these results,
what we show below is that, using a special kind of tuning, 
for predicting the future value of our indicators, we can achieve 
{\bf low predictions error rates} 
(usually under 25\%).
We conjecture that our error rates are so low since we use {\bf arguably better technology} than prior work.
Most of the prior work neglects to tune the control parameters of their learners.
This is not ideal since some recent research in SE reports that such tuning can significantly improve the performance of models used in software analytics~\cite{Tantithamthavorn16,fu2016differential,Fu2016TuningFS,agrawal2018betterdata,agrawal2019dodge,agrawal2018better,9463120, tu2021frugal}.
 Here,  we explore a set of optimization technologies such as Random Search, Grid Search, Differential Evolution, FLASH and auto-sklearn to automatically tune  our learners.
While future work
might discover better optimization methods, what we can say in this paper is that for health indicator predictions, hyperparameter optimization (HPO) can help models to make better predictions.

This paper makes extensive
use of recent data to predict the values of multiple health indicators of open-source projects.
Looking at prior work that studied multiple health indicators,
two closely  comparable studies to this paper are {\em Healthy or not: A way to predict ecosystem health in GitHub} by 
Liao et al.~\cite{liao2019healthy}
and{ \em A Large Scale Study of Long-Time Contributor Prediction for GitHub Projects} by
Bao et al.~\cite{bao2019large}.
These papers studied project developing activities on hundreds of projects. On the basis of their work, we would like to take a further step, to explore more. In our study, we explore 1,159 projects, much of our data is current (as of April 2020) while some prior works use project data that is years to decades old~\cite{sarro2016multi}.


One unique aspect of this work, compared to prior work, is that
we try to explore more kinds of indicators than those seen in prior papers.  For example, the goal of the Bao et al. paper~\cite{bao2019large} is to predict if  a programmer will become a long term  contributor to a GitHub project. While this is certainly   an important question, it is all about {\em individuals} within a single project. The goal of our paper is to offer management advice at a  {\em project level}.

\subsection{Contributions and Structure of  this Paper}

The specific  contributions of this paper are:
\bi
\item Our developer surveys show that the project health issues explored here are related to real-world developer concerns for real projects (see Section~\ref{sect:survey}), to the best of our knowledge, this is the largest developer survey yet explored for open-source project health.
\item We show the viability of large scale project health analysis, for multiple indicators (prior work usually explores just one~\cite{kikas2016using,qi2017software,aggarwal2014co,han2019characterization,weber2014makes} or two~\cite{jarczyk2018surgical,bidoki2018cross} indicators).
\item This is the first paper to demonstrate that project health indicators can be predicted  into the future. Specially, when we compare the errors in our project health estimators 1, 3, 6 and 12 months into the future, we find that~(a)~the error for predicting one month into the future is very low (25\% MRE error, see \tbl{med_mre}) and (b)~that error does not increase much as we look further into the future (less than 30\%, see \tbl{change}).
\item We show the viability
of hyperparameter optimization
across a very large number of projects.
Prior to this work, the experience was that such optimization can be very slow~\cite{Fu2016TuningFS} and hence might not scale to a large population of projects. However, what was realized here is that the optimization problem
for 1,000+ projects is thousand of small problems. Hence there is no need here for elaborate hyperparameter optimization. For example,
in our experiments, we find that a very simple optimizer (differential evolution applied to a CART regression tree) can reach similar prediction performance as a state-of-the-art optimizing system (auto-sklearn~\cite{feurer2019auto}), and does so much  faster.
\ei

More generally, we offer here a somewhat novel  perspective on the process
of hyperparameter optimization and automatic tuning of learners for software analytics.
In our view, it is very important to discuss and reduce the cost of hyperparameter  optimization
since it may not be a ``one-off'' cost.
Rather, it may have to be repeated many times during a project.
Recently we explored an environment where     tunings from one
application were applied to another.   In that work we did the
optimizations again for the new application  and  found that a combination of
using (some) old information plus (new) tuning lead to best results~\cite{9050841}.
While this paper does not explore tuning transfer, our lesson
from that other work is that tuning needs to be specialized whenever the data changes; i.e it should be repeated many times along the life cycle.

But there is a problem:    in many domains,  tuning is CPU-intensive and  
needs   cloud-based CPU  to terminate quickly~\cite{Fu2016TuningFS}.
But just because that is true in many domains, it does
not mean it is true in all domains.
The lesson of this
paper is that 
{\em the tuning of tiny regression
trees on very small
data sets is a specialized domain with its own properties.}
Here, 1,159 times, we run optimizers to better
 extract a signal from data sets with around a dozen
parameters and (at most) 60 rows of data. Hence, as we show below, such optimization
can be complete very quickly (in less than 15 seconds).
Further, at least for the studies we report here, longer optimization (that run 100 times
longer) using more complex optimizers (auto-sklearn), perform no better that our
very fast, very simple optimisers\footnote{In a recent TSE'21 article.
we have explained by SE hyperparameter optimization can be so simple:
SE data can be intrinsically simpler than other kinds of data and, hence
 simpler to explore (see Figure 6d of~\cite{Agrawal21}).}.

The rest of this paper is organized as follows:
Section~\ref{sect:backg} introduces the current problems of open-source software development, the background of software project health, the related work on software analytics of open-source projects, and the difference between our work and prior studies.
After that, Section~\ref{sect:empir} describes the research questions, explain our open-source project data mining, model constructions and the experiment setup details. 
Section~\ref{sect:resul} presents the experimental results and answers the research questions. 
This is followed by Section~\ref{sect:discu} and Section~\ref{sect:threa}, which discuss the findings from the experiment and the potential threats in the study. 
Finally, the conclusions and future works are given in Section~\ref{sect:concl}.

For the health indicators data used in this work,
please see our Github repository\footnote{\url{https://github.com/arennax/Health_Indicator_Prediction}}.

\section{Background and Related Work}
\label{sect:backg}
In this section, we show the background and motivation of our study, current related works from other researchers, and the methods to use in our experiment.

\subsection{Why Study Project Health Indicators?}
\label{sect:survey}

There are many business scenarios within which the predictions of this paper would be very useful.
This section discusses those scenarios. 

We study open source software since it is becoming more prominent in the overall software engineering landscape. As said in our introduction, over 80\% of the software in any technology product or service is now open-source~\cite{zemlin2017}. 

As the open source community matures, so too does its management practices.
Increasingly, open source projects are becoming more structured with organizing foundations. For example, as the largest software organizations, the Apache Software Foundation and Linux Foundation currently host 371 projects and 166 projects respectively~\cite{apacheprojects,linuxprojects}. 
For the promising projects, those organizations invest significant funds to secure  seats on the board of those projects.
Although the stakeholders of these projects have different opinions, they all need indicators of project status to make decisions. For example, project managers need information about upcoming activities in the development to decide where to put resources, and justifications to convince that it is cost effective to pay developers to work on specific tasks. 

Some engineers might be  eager to attract funding to the foundations that run large open source projects. Large organizations are willing to pay for the privilege of participating in the governance of projects. Hence, it is important to have a good ``public profile'' for the projects to keep these organizations interested. In the case of GitHub projects, the feature ``number of stars'' has been recognized as a success and popularity measurement to look at~\cite{han2019characterization,borges2016understanding,borges2016predicting}. 

Also, many developers must regularly report the status of their projects to the governing foundations. Those reports will determine the allocations of future resources for these projects. For such reports, it is important to assess the projects compared to other similar projects.
For example, some program managers argue that their project is scoring ``better'' than other similar projects since that project has more new contributors per month~\cite{xia21}. In our study, the feature ``number of contributors'' is a suitable measure for this way of scoring.

We also note that 
many commercial companies   use open-source packages in the products they sell to customers. For that purpose, commercial companies want to use relatively stable packages (i.e. no massive abnormal developing activities) for some time to come. Otherwise, if the open-source community stops maintaining or changes those
packages, then those companies
will be forced into maintaining open-source packages which they may not fully understand.

Another case where commercial organizations can use predictions of project health indicators is the issue
of {\em ecosystem package management}.
For example, Red Hat are very interested in project health indicators that can be automatically applied to tens of thousands of projects. 
 When they release a new version of their OS, more than 24,000 software packages included in the distribution are   delivered to tens of millions of machines around the world.  For this process, Red Hat seek project health indicators to help them:
\bi
\item Decide what packages should not be included in the next distribution (due to abnormal behaviors in the development);
\item Detect, then repair, falling health in   popular packages.
For example, in 2019, Red Hat's engineers
noted that a particularly popular  project was falling out of favor with other developers since its regression test suite was not keeping up with current changes. With just a few thousand dollars, Red Hat used crowd sourced programmers
to generate the tests that made the package viable again~\cite{stewart19}.
\ei 

Yet another use case where predictions of project health indicators would be useful is {\em software staff management}.
  Thousands of IBM developers maintain dozens of    large open-source toolkits. 
 IBM needs to know the expected workload within those projects, several months in advance~\cite{krishna2018connection}.
Some indicators can advise when  there are too many developers working on one project, and not enough working on another.
Using this information,
 IBM management  can    ``juggle'' that staff around multiple  projects in order to match the  expected workload to the available staff. 
 For example, 
 \bi
 \item
 If a spike is expected in a few months for the number of pull
 requests,    management might move extra staff over to that project a couple of months earlier (so that staff can learn that codebase). 
 \item When handling  the training of newcomers, it is unwise to drop novices into some high stress scenarios where too few programmers are struggling to handle a large workload.
 \item It is also useful to know when the workload for a project is predicted to be stable or decreasing. In that use case, it is not ill-advised to move staff to other problems in order to accommodate the requests of seasoned programmers who want to either learn new technologies as part of their career development; or resolve personnel conflict issues.
 \ei

In our study, we focus on the 7 critical and easy to access GitHub features listed in Section~\ref{sect:intro} as our potential health indicators.

\begin{wrapfigure}{r}{2.5in}
\centering
\scriptsize 
\vspace{-0.8cm}
\begin{center}
\begin{lstlisting}[language=bash,linewidth=6cm,frame=single,numbers=none,keywords=none]
Question 1
Are you a core contributor to this project? 
(Yes / No)

Question 2
Did you notice your project was having a 
health problem in the previous development? 
(Yes / No)

Question 3
Please rate the importance of the following 
features regarding project health: 
(1:Very Important
 2:Somewhat Important
 3:Not Important)
 
 # contributors,   # commits,   # opened pull-requests, 
 # closed pull-requests,   # opened issues, 
 # closed issues, # stargazers

Question 4
Besides the features mentioned in last question,
what else developing features do you think are 
relevant to open source project health?
\end{lstlisting} 
\end{center}
\vspace{-0.5cm}
\caption{Transcript of survey questions.}
\label{fig:transcript} 
\vspace{-0.5cm}
\end{wrapfigure}

To verify these features actually matter in real-world business-level cases, we have conducted extensive interviews with real-world open source software developers. From January to March, 2021, we selected 100 open source projects from our data collection list, and sent email surveys to their main developers to ask about their opinions on health indicators based on their development experience. We made the survey questions short and easy to reply to in order to get more responses and higher engagements. The transcript of our survey questions is  shown in \fig{transcript}.
In the end, 112 core contributors
from 68 projects provide valid responses. Table~\ref{tbl:survey} shows the summary of whether our indicators matter to project health in their opinions.

Based on Table~\ref{tbl:survey}, we can say features like ``contributors'', ``commits'', ``opened pull-request'', ``closed pull-request'', ``opened issues'' and ``closed issues'' are all important as health indicators in the survey since very few developers treat them ``Not Important''(mostly less than 20\%). The only exception is ``stargazers'', which the majority of the participants (52\%) consider as ``Not Important''. Hence, we conclude the first 6 features have great impacts on project health based on experienced software engineers' judgements and could be used as our health indicators in the experiment. 

\begin{table}[!t]
\centering
\caption{Importance of 7 Indicators to Project Health (based on the survey).}
\label{tbl:survey}
\begin{adjustbox}{max width=0.98\textwidth}
\begin{tabular}{l|c|c|c}
\rowcolor[HTML]{BDBDBD} 
 & Very Important & Somewhat Important & Not Important \\ \hline
\rowcolor[HTML]{FFFFFF} 
contributors & 33\% & 46\% & 21\% \\
\rowcolor[HTML]{F3F3F3} 
commits & 57\% & 31\% & 12\% \\
\rowcolor[HTML]{FFFFFF} 
opened pull-requests & 31\% & 56\% & 13\% \\
\rowcolor[HTML]{F3F3F3} 
closed pull-requests & 46\% & 34\% & 20\% \\
\rowcolor[HTML]{FFFFFF} 
opened issues & 33\% & 59\% & 8\% \\
\rowcolor[HTML]{F3F3F3} 
closed issues & 40\% & 53\% & 7\% \\
\rowcolor[HTML]{FFFFFF} 
stargazers & 12\% & 36\% & 52\%
\end{tabular}
\end{adjustbox}
\end{table}
 

\subsection{How to Study Project Health Indicators?}
\label{tion:lit}
The results of our interviews and surveys give us hints that some activities in open source development can be treated as indicators of project health. In our literature review, we find numerous studies and organizations are exploring the health or development features of open-source projects. 
For example:
\bi
\item
Jansen et al.~\cite{jansen2014measuring} introduce an OSEHO (Open Source Ecosystem Health Operationalization) framework, using productivity, robustness and niche creation to measure the health of software ecosystem.
\item
Manikas et al.~\cite{manikas2013reviewing} propose a logical framework for defining and measuring the software ecosystem health consisting of the health of three main components (actors, software and  orchestration).
\item
A community named ``CHAOSS'' (Community Health Analytics for Open Source Software)~\cite{chaoss} contributes on developing metrics, methodologies, and software from a wide range of open-source projects to express open-source project health and sustainability.
\item
Weber et al.~\cite{weber2014makes} use a random forest classifier to predict project popularity (which they define as  the star velocity in their study) on a set of Python projects.
\item
Borges et al.~\cite{borges2016predicting} claim that the number of stars of a repository is a direct measure of its popularity, in their study, they use a model with multiple linear regressions to predict the number of stars to estimate the popularity of GitHub repositories.

\item
Wang et al.~\cite{wang2018will} propose a prediction model using regression analysis to find potential long-term contributors (through their capacity, willingness, and the opportunity to contribute at the time of joining). They validate their methods on ``Ruby on Rails'', on a large and popular project on GitHub. 
Bao et al.~\cite{bao2019large} use a set of methods (Naive Bayes, SVR, Decision Tree, KNN and Random Forest) on 917 projects from GHTorrent to predict long term contributors (which they determine as the time interval between their first and last commit in the project is larger than a threshold.), they create a benchmark for the result and find random forest achieves the best performance.
\item
Kikas et al.~\cite{kikas2016using} build random forest models to predict the issue close time on GitHub projects, with multiple static, dynamic and contextual features. They report that the dynamic and contextual features are critical in such
prediction tasks.
Jarczyk et al.~\cite{jarczyk2018surgical} use generalized linear models for prediction of issue closure rate. Based on multiple features (stars, commits, issues closed by team, etc.), they find that larger teams with more project members have lower issue closure rates than smaller teams, while increased work centralization improves issue closure rates. 
\item
Other developing related feature predictions also include the information of commits, which is used by Qi et al.~\cite{qi2017software} in their software effort estimation research of open source projects, where they treat the number of commits as an indicator of human effort.
\item
Chen et al.~\cite{chen2014predicting} use linear regression models on 1,000 GitHub projects to predict the number of forks, they conclude this prediction could help GitHub to recommend popular projects, and guide developers to find projects which are likely to succeed and worthy of their contribution.

\ei

We explore the literature looking for how prior researchers have explored software developing activities. 
Starting with venues listed at Google Scholar Metrics 
``software systems'', 
we searched for highly cited or very recent papers discussing 
{\em software analytics, project health, open  source systems} and {\em GitHub predicting}.
We found:
\bi
\item In the past six years (2014 to 2020), there were at least 30 related papers.
\item 10 of those papers looked at least one of the seven features we listed in our introduction~\cite{liao2019healthy,borges2016predicting,jarczyk2018surgical,kikas2016using,qi2017software,aggarwal2014co,chen2014predicting,han2019characterization,weber2014makes,bidoki2018cross}.
\item 3 of those papers explored multiple features~\cite{liao2019healthy,jarczyk2018surgical,bidoki2018cross}.
\ei

Following the previous research, we consider these features as project health indicators, make a massive, systematic time-series data collection, and try to find a general method to predict the trends of these indicators.

As to the technology used in the related papers, the preferred predicting method was usually just one of the following:

\bi
\item  LNR: {\em linear regression} model that builds regression methods to fit the data to a parametric equation; 
\item CART: {\em decision tree learner} for classification and regression;
\item RFT:  {\em random forest}   that builds multiple regression trees, then report the average conclusion across that forest;
\item KNN:  {\em k-nearest neighbors} that make conclusions by average across nearby examples;
\item SVR:   {\em support vector regression} uses the regressions that take the quadratic optimizer used in support vector machines and use it to learn a parametric equation that predicts for a numeric class. 
\ei

Hence, for this study, we use the above learners as baseline methods with implementations from Scikit-Learn~\cite{pedregosa2011scikit}. Unless being adjusted by hyperparameter optimizers (discussed below), all these learners run with the default settings.\footnote{
We use default settings for the baselines to find if they can provide good prediction performance, and how much space hyperparameter-tuning can improve. Using a pre-selected parameter-settings from literature may bring bias because of different data format or prediction tasks.
}

Of the above related work, a study by 
Bao et al. from TSE'19 seems close to our work~\cite{bao2019large}. 
They explored multiple learning methods for their prediction tasks.
Further, while the other papers used learners with their off-the-shelf settings, Bao et al. took care to tune the control
hyperparameters of their learners.


The idea of tuning the control hyperparameters to improve the prediction performance has been applied to many machine learning algorithms. For example, Bergstra et al. used random search (i.e. just pick
parameters at random)
and greedy sequential methods on finding the best configurations of their neural networks and deep belief networks models~\cite{bergstra2011algorithms}. Snoek et al. applied Bayesian optimization to reduce the prediction errors of logistic regression and support vector machines~\cite{snoek2012practical}.

Much recent research in SE report that such hyperparameter tuning can significantly improve the performance of prediction methods used in many software  analytic tasks~\cite{fu2016differential,Fu2016TuningFS,agrawal2018betterdata,agrawal2019dodge,agrawal2018better,xia2020sequential}. For example, for defect prediction, 
Fu et al. applied an evolutionary algorithm named Differential Evolution (DE)~\cite{storn1997differential} on a set of tree-based defect predictors (e.g. CART). With the data from open source JAVA systems, their experiment results showed that hyperparameter tuning largely improves the precision of these predictors~\cite{Fu2016TuningFS}.
Agrawal et al. also used Differential Evolution to automatically tune hyperparameters of a processor named ``SMOTE''. Based on the results of seven datasets, their experiments showed that Differential Evolution can lead to up to 60\% improvements in predictive performance when tuning SMOTE’s hyperparameters~\cite{agrawal2018betterdata}.
Tantithamthavorn et al. investigated the impact of hyperparameter tuning on a case study with 18 datasets and 26 learning models. They tuned multiple hyperparameters using a grid search (looping over all possible parameter settings) on 100 repetitions of the out-of-sample bootstrap procedure. With that approach, they found results led to improvements of up to 40\% in the Area Under the receiver operator characteristic Curve~\cite{Tantithamthavorn16}.

In software system configurations, Nair et al. proposed FLASH, a sequential model-based optimizer, which explored the configuration space by reflecting on the configurations evaluated so far to find the best configuration for the systems, they evaluated FLASH on 7 software systems and demonstrated that FLASH can effectively find the best configuration~\cite{nair2017flash}. Later, Xia et al. applied FLASH to tune the hyperparameters of CART in their study of software effort estimation, using data from 1,161 waterfall projects and 120 contemporary projects, the results showed that this sequential model-based optimization achieved better performance than previous state-of-the-art methods~\cite{xia2020sequential}. Also, in effort estimation, Minku et al. proposed an online supervised hyperparameter tuning procedure, which helps to tune the number of clusters in Dycom (Dynamic Cross-company Mapped Model Learning), a software effort estimation online ensemble learning approach. Using the ISBSG Repository, they showed that the proposed method was generally successful in enabling a very simple threshold-based clustering approach to obtain the most competitive Dycom results~\cite{minku2019novel}.
In this study, we do not use methods of Minku et al. since,
at least so far, our work has been on within-company estimation (i.e. learning from the history of
some {\em current project}, then applying what was learned to later
points in that same project).

Recently, Tantithamthavorn et al.~\cite{tantithamthavorn2018impact}
extended their defect prediction study on more hyperparameter optimizers with 
genetic algorithms, random search, 
and Differential Evolution.
They found that in the defect prediction domain, different hyperparameter tuning procedures led to similar benefits in terms of performance improvement; i.e. at least in that domain, it may not be necessary to perform extensive studies of across different hyperparameter optimizers.~\cite{tantithamthavorn2018impact}.

For this paper, echoing the methods of the advice of Tantithamthavorn et al., we explore  hyperparameter optimization using Grid Search, Random Search, FLASH and Differential Evolution. We also add a state-of-the-art HPO technology ``auto-sklearn'' into our experiments to verify the perofrmance of hyperparameter optimization. ``auto-sklearn'' is an automated machine learning toolkit and a drop-in replacement for a scikit-learn estimator, it automatically searches for the right learning algorithm for a new machine learning dataset and optimizes its hyperparameters~\cite{feurer2019auto}.
In our experiments, we will show that methods with hyperparameter optimization get better results than untuned methods. In this regard, we offer the verification as other researchers who have found HPO to be useful for tuning  software analytic problems (e.g., defect prediction~\cite{fu2016differential,Fu2016TuningFS}). 
Also, since some HPO methods (e.g. DE) has shown better performance comparing other methods like random search, further exploration of other algorithms as hyperparameter optimizers may bring even better performance (this could be part of the future work).

\begin{figure}[!b]
\centering
\small 
\begin{center}
\begin{minipage}{4in}\begin{lstlisting}[mathescape,linewidth=7.5cm,frame=none,numbers=left ]
  def DE(np=20, cf=0.75, f=0.3, lives=10):  # default settings
    frontier = # make "np" number of random guesses
    best = frontier.1 # any value at all
    while(lives$--$ > 0): 
      tmp = empty
      for i = 1 to $|$frontier$|$: # size of frontier
         old = frontier$_i$
         x,y,z = any three from frontier, picked at random
         new= copy(old)
         for j = 1 to $|$new$|$: # for all attributes
          if rand() < cf    # at probability cf...
              new.j = $x.j + f*(z.j - y.j)$  # ...change item j
         # end for
         new  = new if better(new,old) else old
         tmp$_i$ = new 
         if better(new,best) then
            best = new
            lives++ # enable one more generation
         end
      # end for
     frontier = tmp
     lives--
    # end while
    return best
\end{lstlisting} 
\end{minipage}
\end{center}
\caption{Differential evolution. Pseudocode based on Storn's algorithm~\cite{storn1997differential}.}
\label{fig:pseudo_DE} 
\vspace{-0.3cm}
\end{figure}

The pseudocode of DE algorithm is shown in \fig{pseudo_DE}. The premise of that code is that the best way to mutate the existing tunings is to extrapolate between current solutions (stored in the {\em frontier} list). Three solutions $x, y, z$ are selected at random from the {\em frontier}. For each tuning parameter $j$, at some probability $cf$ (crossover probability), DE  replaces the old tuning $x_j$ with {\em new}  where
\mbox{$\mathit{new}_j = x_j + f \times (y_j - z_j)$}
where $f$ (differential weight) is a parameter controlling differential weight.

The main loop of DE runs over the {\em frontier} of size $np$ (population size), replacing old items with new candidates (if new candidate is better). This means that, as the loop progresses, the {\em frontier}  contains increasingly more valuable solutions (which, in turn, helps extrapolation since the next time we pick $x,y,z$, we get better candidates.). 

DE's loops keep repeating till it runs out of {\em lives}. The number of {\em lives} is decremented for each loop (and incremented every time we find a better solution).

Our initial  experiments 
showed that out of all these ``off-the-shelf'' learners, the CART regression tree learner was performing best. Hence, we combine CART with differential evolution to create the DECART hyperparameter  optimizer for CART regression trees. 
The choice of these parameters can have a large impact on optimization performance. Taking advice from  Storn and Fu et al.~\cite{storn1997differential,Fu2016TuningFS}, we set DE's configuration parameters to $\{\mathit{np, cf, f, \mathit{lives}}\}=\{20,0.75,0.3,10\}$. The CART hyperparameters we control via DE are shown in Table~\ref{tbl:cart}. For the tuning ranges of each hyperparameters, we follow the suggestions from Fu et al.~\cite{Fu2016TuningFS}.

\begin{table}[!h]
\centering
\caption{The hyperparameters to be tuned in CART.}
\label{tbl:cart}
\begin{adjustbox}{max width=0.98\textwidth}
\begin{tabular}{l|c|c|l}
\rowcolor[HTML]{BDBDBD} 
{\color[HTML]{000000} Hyperparameter} & \multicolumn{1}{l|}{\cellcolor[HTML]{BDBDBD}{\color[HTML]{000000} Default}} & \multicolumn{1}{l|}{\cellcolor[HTML]{BDBDBD}{\color[HTML]{000000} Tuning Range}} & {\color[HTML]{000000} Description} \\ \hline
\rowcolor[HTML]{FFFFFF} 
{\color[HTML]{000000} max\_feature} & {\color[HTML]{000000} None} & {\color[HTML]{000000} {[}0.01, 1{]}} & {\color[HTML]{000000} \begin{tabular}[c]{@{}l@{}}Number of features to consider \\ when looking for the best split\end{tabular}} \\
\rowcolor[HTML]{F3F3F3} 
{\color[HTML]{000000} max\_depth} & {\color[HTML]{000000} None} & {\color[HTML]{000000} {[}1, 12{]}} & {\color[HTML]{000000} \begin{tabular}[c]{@{}l@{}}The maximum depth of the \\ decision tree\end{tabular}} \\
\rowcolor[HTML]{FFFFFF} 
{\color[HTML]{000000} min\_sample\_leaf} & {\color[HTML]{000000} 1} & {\color[HTML]{000000} {[}1, 12{]}} & {\color[HTML]{000000} \begin{tabular}[c]{@{}l@{}}Minimum samples required to \\ be at a leaf node\end{tabular}} \\
\rowcolor[HTML]{F3F3F3} 
{\color[HTML]{000000} min\_sample\_split} & {\color[HTML]{000000} 2} & {\color[HTML]{000000} {[}0, 20{]}} & {\color[HTML]{000000} \begin{tabular}[c]{@{}l@{}}Minimum samples required to \\ split internal nodes\end{tabular}}
\end{tabular}
\end{adjustbox}
\end{table}

\section{Experiment Setup}
\label{sect:empir} 
In this section, we ask the related research questions, provide details of our experiment data, explain how to construct our experiment model, and how to measure our experiment results.

\subsection{Research Questions}
The experiments of this paper are designed to explore the following questions.

\textbf{RQ1: How to find the trends of project health indicators?}
We apply five popular machine learning algorithms (i.e., KNN, SVR, LNR, RFT and CART), and five hyperparameter-optimized methods: CART tuned by random search (RDCART), grid search (GSCART), FLASH, differential evolution (DECART) and auto-sklearn (ASKL) with the same trial budget\footnote{i.e. A maximum of 200 evaluations for Random Search, Grid Search, Flash and  DE; for ASKL, maximum runtime for each project is restricted to 15 seconds, please see Section~\ref{sect:runtime} for details.} to 1,159 open-source projects collected from GitHub (For FLASH, we apply the same settings as used in the previous work~\cite{xia2020sequential}). 
For each project, once we collect $N$ months of data, we make predictions for the recent status using a part of data from month 1 to month $\mathit{N-j}$ (for $\mathit{j\in\{1,3,6,12\}}$) in the past.
In the experiments, the median prediction errors of health indicators can be reached under 25\% (where  this error is  calculated from $\mathit{error}=100*|p-a|/a$  using the predicted $p$ and actual $a$).
Hence, we  will say:




\begin{blockquote}
\noindent
\textbf{Answer 1}: Many project health indicators can be predicted, with good accuracy, for 1, 3, 6, 12   months into the future.
\end{blockquote} 

\textbf{RQ2: What features matter the most in prediction?}
To find the most important features that have been used for prediction, we look into the internal structure of model with the best prediction and compute impurity-based feature importances (Gini importance). We will show that:

\begin{blockquote}
\noindent
\textbf{Answer 2}: In our study, ``monthly\_commits'', ``monthly\_openPR'', ``monthly\_openISSUE'' and ``monthly\_closeISSUE'' are the most important features, while ``monthly\_closePR'' is the least used feature for all six health indicators' predictions.
\end{blockquote}

\textbf{RQ3: How to improve the performance of health indicators prediction?}
We compare the experimental results of each method on all 1,159 open-source projects and prediction for 1, 3, 6, and 12 months into the future. After a statistical comparison between different learners, we find that:  

\begin{blockquote}
\noindent
\textbf{Answer 3}: Hyperparameter optimized methods generate better prediction performance than the other methods.
\end{blockquote}


\subsection{Data Collection}
\label{sect:data_collect}

Many repositories on GitHub are not suitable for software engineering research~\cite{kalliamvakou2016depth,munaiah2017curating}. We follow advice from Kalliamvakou et al. and Munaiah et al., apply related criteria (with GitHub GraphQL API) to find useful URLs of the projects~\cite{munaiah2017curating,kalliamvakou2014promises}.
As shown in Table~\ref{tbl:select}, we select public, not archived, and not mirrored repositories as open sources, use a set of thresholds to ensure they are active and collaborative, with relatively popular profiles (Shrikanth et al. report that popular projects usually have stars around 1k to 20k~\cite{shrikanth2021early}).

In addition, to remove repositories with irrelevant topics such as ``books'', ``class projects'' or ``tutorial docs'', etc., we create a dictionary of ``suspicious words of irrelevancy'', and remove URLs that contain words in that dictionary (see  Table~\ref{tbl:dict}). After applying the criteria of Table~\ref{tbl:select}, Table~\ref{tbl:dict} and a round of manual checking, 
we get 1,159 repositories which we treat as engineered software projects. From these repositories, we extract features of in total 64,181 monthly data across all projects.

At the point of writing, there is no unique and consolidated definition of software project health~\cite{jansen2014measuring,liao2019healthy,link2018assessing}. However, many researchers agree that healthy open-source projects need to be ``vigorous'' and ``active''~\cite{wahyudin2007monitoring,jansen2014measuring,manikas2013reviewing,link2018assessing,wynn2007assessing,crowston2006assessing}. 
Based on our previous survey, we select 6 features as health indicators of open-source project on GitHub: number of commits, contributors, open pull-requests, closed pull-requests, open issues and closed issues. These features are important GitHub features to indicate the activities of the projects~\cite{borges2016understanding,han2019characterization,aggarwal2014co}.

All the features collected from each project in this study 
are listed in Table~\ref{tbl:feature}. These features are carefully selected because some of them are used by other researchers who explore related
GitHub studies~\cite{coelho2020github, yu2016reviewer, han2019characterization}.  

\begin{table} 
\centering
\caption{Repository selecting criteria.}
\label{tbl:select}
\begin{adjustbox}{max width=0.98\textwidth}
\begin{tabular}{l|l}
\rowcolor[HTML]{BDBDBD} 
{\color[HTML]{000000} Filter} & {\color[HTML]{000000} Explanation} \\ \hline
\rowcolor[HTML]{FFFFFF} 
{\color[HTML]{000000} is:public} & {\color[HTML]{000000} select open-source repo} \\
\rowcolor[HTML]{F3F3F3} 
{\color[HTML]{000000} archived:false} & {\color[HTML]{000000} exclude archived repo} \\
\rowcolor[HTML]{FFFFFF} 
{\color[HTML]{000000} mirror:false} & {\color[HTML]{000000} exclude duplicate repo} \\
\rowcolor[HTML]{F3F3F3} 
{\color[HTML]{000000} stars:1000..20000} & {\color[HTML]{000000} select relatively popular repo} \\
\rowcolor[HTML]{FFFFFF} 
{\color[HTML]{000000} size:\textgreater{}=10000} & {\color[HTML]{000000} exclude too small repo} \\
\rowcolor[HTML]{F3F3F3} 
{\color[HTML]{000000} forks:\textgreater{}=10} & {\color[HTML]{000000} select repo being forked} \\
\rowcolor[HTML]{FFFFFF} 
{\color[HTML]{000000} created:\textgreater{}=2015-01-01} & {\color[HTML]{000000} select relatively new repo} \\
\rowcolor[HTML]{F3F3F3} 
{\color[HTML]{000000} created:\textless{}=2016-12-31} & {\color[HTML]{000000} select repo with enough monthly data} \\
\rowcolor[HTML]{FFFFFF} 
{\color[HTML]{000000} contributor:\textgreater{}=3} & {\color[HTML]{000000} exclude personal repo} \\
\rowcolor[HTML]{F3F3F3} 
{\color[HTML]{000000} total\_commit:\textgreater{}=1000} & {\color[HTML]{000000} select repo with enough commits} \\
\rowcolor[HTML]{FFFFFF} 
{\color[HTML]{000000} total\_issue\_closed:\textgreater{}=50} & {\color[HTML]{000000} select repos with enough issues} \\
\rowcolor[HTML]{F3F3F3} 
{\color[HTML]{000000} total\_PR\_closed:\textgreater{}=50} & {\color[HTML]{000000} select repos with enough pull-request} \\
\rowcolor[HTML]{FFFFFF} 
{\color[HTML]{000000} recent\_PR:\textgreater{}=1 (30 days)} & {\color[HTML]{000000} exclude inactive repo without PR} \\
\rowcolor[HTML]{F3F3F3} 
{\color[HTML]{000000} recent\_commit:\textgreater{}=1 (30 days)} & {\color[HTML]{000000} exclude inactive repo without commits}
\end{tabular}
\end{adjustbox}
\end{table}

\begin{table}[!b] 
\centering
\begin{adjustbox}{max width=0.98\textwidth}
\begin{tabular}{ccccccc}
\rowcolor[HTML]{BDBDBD} 
\multicolumn{7}{c}{\cellcolor[HTML]{BDBDBD}{\color[HTML]{000000} Suspicious Keywords}} \\ \hline
\rowcolor[HTML]{FFFFFF} 
{\color[HTML]{000000} template} & {\color[HTML]{000000} web} & {\color[HTML]{000000} tutorial} & {\color[HTML]{000000} lecture} & {\color[HTML]{000000} sample} & {\color[HTML]{000000} note} & {\color[HTML]{000000} sheet} \\
\rowcolor[HTML]{F3F3F3} 
{\color[HTML]{000000} book} & {\color[HTML]{000000} doc} & {\color[HTML]{000000} image} & {\color[HTML]{000000} video} & {\color[HTML]{000000} demo} & {\color[HTML]{000000} conf} & {\color[HTML]{000000} intro} \\
\rowcolor[HTML]{FFFFFF} 
{\color[HTML]{000000} class} & {\color[HTML]{000000} exam} & {\color[HTML]{000000} study} & {\color[HTML]{000000} material} & {\color[HTML]{000000} test} & {\color[HTML]{000000} exercise} & {\color[HTML]{000000} resource} \\
\rowcolor[HTML]{F3F3F3} 
{\color[HTML]{000000} article} & {\color[HTML]{000000} academic} & {\color[HTML]{000000} result} & {\color[HTML]{000000} output} & {\color[HTML]{000000} resume} & {\color[HTML]{000000} work} & {\color[HTML]{000000} guide} \\
\rowcolor[HTML]{FFFFFF} 
{\color[HTML]{000000} present} & {\color[HTML]{000000} slide} & {\color[HTML]{000000} 101} & {\color[HTML]{000000} qa} & {\color[HTML]{000000} view} & {\color[HTML]{000000} form} & {\color[HTML]{000000} course} \\
\rowcolor[HTML]{F3F3F3} 
{\color[HTML]{000000} thesis} & {\color[HTML]{000000} collect} & {\color[HTML]{000000} pdf} & {\color[HTML]{000000} wiki} & {\color[HTML]{000000} blog} & {\color[HTML]{000000} lesson} & {\color[HTML]{000000} pic} \\
\rowcolor[HTML]{FFFFFF} 
{\color[HTML]{000000} paper} & {\color[HTML]{000000} camp} & {\color[HTML]{000000} summit} & {\color[HTML]{000000} } & {\color[HTML]{000000} } & {\color[HTML]{000000} } & {\color[HTML]{000000} }
\end{tabular}
\end{adjustbox}
\caption{
Dictionary of ``irrelevant'' words. We do not use data from projects whose URL includes the following keywords.
}
\label{tbl:dict}
\end{table}


To get the latest and accurate features
of our selected repositories,  we use the GitHub APIs for feature collection. For each project, the first commit date is
used as the starting date of the project. Then all the features are collected and calculated monthly
from that date up to the present date. For example, the first commit of
the {\it kotlin-native} project was on May 16,  2016.
After, we collected  features from May, 2016 to April, 2020. Due to the GitHub API rate limit, we could not get some features, like ``monthly\_commits'', which need many   direct API calls. Instead, we clone the repo locally and then extracted   features (this technique saved us much grief with API  quotas). 
Table~\ref{tbl:data} shows a summary of the data collected by using this method. 

\subsection{Model Construction}
In our experiment, we use five classical machine learning algorithms and five hyperparameter tuned methods for the prediction tasks. These five classical machine learning algorithms are Nearest Neighbors, Support Vector Regression, Linear Regression, Random Forest and Regression Tree (we call them KNN, SVR, LNR, RFT and CART). The configuration of those baselines follows the suggestions from Scikit-Learn.

Beyond the baseline methods, we build hyperparameter optimized predictors, named ``DECART'', ``RDCART'', ``GSCART'' and ``FLASH''. ``DECART'' uses differential evolution algorithm (with configuration settings to {np=20, cf=0.75, f=0.3, lives=10}) as an optimizer to optimize four hyperparameters (max\_feature, max\_depth, min\_sample\_leaf and min\_sample\_split) of regression tree (CART), and use this tuned CART to get predict results.

\begin{table}[!t]
\centering
\caption{Project health indicators.   ``PR''= pull requests.  When predicting feature ``X'' (e.g. \# of commits), we re-arrange the data such the dependent variable is ``X'' and the independent variables are the rest.
}
\label{tbl:feature}
\begin{adjustbox}{max width=0.98\textwidth}
\footnotesize
\begin{tabular}{lllc}
\rowcolor[HTML]{CCCCCC} 
{\color[HTML]{000000} Dimension} & {\color[HTML]{000000} Feature} & {\color[HTML]{000000} Description} & \multicolumn{1}{l}{\cellcolor[HTML]{CCCCCC}{\color[HTML]{000000} Predict?}} \\ \hline
{\color[HTML]{000000} Commits} & {\color[HTML]{000000} \# of commits} & {\color[HTML]{000000} monthly number of commits} & {\color[HTML]{000000} \cmark} \\ \hline
{\color[HTML]{000000} } & {\color[HTML]{000000} \# of open PRs} & {\color[HTML]{000000} monthly number of open PRs} & {\color[HTML]{000000} \cmark} \\
{\color[HTML]{000000} } & {\color[HTML]{000000} \# of closed PRs} & {\color[HTML]{000000} monthly number of closed PRs} & {\color[HTML]{000000} \cmark} \\
\multirow{-3}{*}{{\color[HTML]{000000} Pull Requests}} & {\color[HTML]{000000} \# of merged PRs} & {\color[HTML]{000000} monthly number of merged PRs} & {\color[HTML]{000000} } \\\hline
{\color[HTML]{000000} } & {\color[HTML]{000000} \# of open issues} & {\color[HTML]{000000} monthly number of open issues} & {\color[HTML]{000000} \cmark} \\
{\color[HTML]{000000} } & {\color[HTML]{000000} \# of closed issues} & {\color[HTML]{000000} monthly number of closed issues} & {\color[HTML]{000000} \cmark} \\
\multirow{-3}{*}{{\color[HTML]{000000} Issues}} & {\color[HTML]{000000} \# of issue comments} & {\color[HTML]{000000} monthly number of issue comments} & {\color[HTML]{000000} } \\ \hline
{\color[HTML]{000000} } & {\color[HTML]{000000} \# of contributors} & {\color[HTML]{000000} monthly number of active contributors} & {\color[HTML]{000000} \cmark} \\
{\color[HTML]{000000} } & {\color[HTML]{000000} \# of stargazers} & {\color[HTML]{000000} monthly increased number of stars} & {\color[HTML]{000000} } \\
\multirow{-3}{*}{{\color[HTML]{000000} Project}} & {\color[HTML]{000000} \# of forks} & {\color[HTML]{000000} monthly increased number of forks} & {\color[HTML]{000000} }
\end{tabular}
\end{adjustbox}
\end{table}

\begin{table}[!b]
\centering
\begin{adjustbox}{max width=0.98\textwidth}
\begin{tabular}{l|cccc}
\rowcolor[HTML]{BDBDBD} 
{\color[HTML]{000000} Feature} & \multicolumn{1}{c}{\cellcolor[HTML]{BDBDBD}{\color[HTML]{000000} Min}} & \multicolumn{1}{c}{\cellcolor[HTML]{BDBDBD}{\color[HTML]{000000} Max}} & \multicolumn{1}{c}{\cellcolor[HTML]{BDBDBD}{\color[HTML]{000000} Median}} & \multicolumn{1}{c}{\cellcolor[HTML]{BDBDBD}{\color[HTML]{000000} IQR}} \\\hline
\rowcolor[HTML]{FFFFFF} 
{\color[HTML]{000000} monthly commits} & {\color[HTML]{000000} 0} & {\color[HTML]{000000} 10358} & {\color[HTML]{000000} 45} & {\color[HTML]{000000} 83} \\
\rowcolor[HTML]{F3F3F3} 
{\color[HTML]{000000} monthly contributors} & {\color[HTML]{000000} 0} & {\color[HTML]{000000} 312} & {\color[HTML]{000000} 6} & {\color[HTML]{000000} 7} \\
\rowcolor[HTML]{FFFFFF} 
{\color[HTML]{000000} monthly stars} & {\color[HTML]{000000} 0} & {\color[HTML]{000000} 6085} & {\color[HTML]{000000} 32} & {\color[HTML]{000000} 68} \\
\rowcolor[HTML]{F3F3F3} 
{\color[HTML]{000000} monthly opened PRs} & {\color[HTML]{000000} 0} & {\color[HTML]{000000} 3860} & {\color[HTML]{000000} 6} & {\color[HTML]{000000} 22} \\
\rowcolor[HTML]{FFFFFF} 
{\color[HTML]{000000} monthly closed PRs} & {\color[HTML]{000000} 0} & {\color[HTML]{000000} 14699} & {\color[HTML]{000000} 1} & {\color[HTML]{000000} 3} \\
\rowcolor[HTML]{F3F3F3} 
{\color[HTML]{000000} monthly merged PRs} & {\color[HTML]{000000} 0} & {\color[HTML]{000000} 1418} & {\color[HTML]{000000} 4} & {\color[HTML]{000000} 18} \\
\rowcolor[HTML]{FFFFFF} 
{\color[HTML]{000000} monthly open issues} & {\color[HTML]{000000} 0} & {\color[HTML]{000000} 3883} & {\color[HTML]{000000} 28} & {\color[HTML]{000000} 53} \\
\rowcolor[HTML]{F3F3F3} 
{\color[HTML]{000000} monthly closed issues} & {\color[HTML]{000000} 0} & {\color[HTML]{000000} 20376} & {\color[HTML]{000000} 24} & {\color[HTML]{000000} 50} \\
\rowcolor[HTML]{FFFFFF} 
{\color[HTML]{000000} monthly issue comments} & {\color[HTML]{000000} 0} & {\color[HTML]{000000} 97846} & {\color[HTML]{000000} 134} & {\color[HTML]{000000} 309} \\
\rowcolor[HTML]{F3F3F3} 
{\color[HTML]{000000} monthly forks} & {\color[HTML]{000000} 0} & {\color[HTML]{000000} 2789} & {\color[HTML]{000000} 7} & {\color[HTML]{000000} 13}
\end{tabular}
\end{adjustbox}
\caption{ 
Summary of 64,181 monthly data across all 1,159 projects.
}
\label{tbl:data}
\end{table}

 We also add a state-of-the-art HPO system named ``auto-sklearn'' into our experiments to compare with our hyperparameter optimized predictors. To make a fair comparison, for each project, the runtime budget of ``auto-sklearn'' is restricted to 15 seconds (which is the upper bound of mean runtime of other hyperparameter optimized predictors), please see Section~\ref{sect:runtime} for details.

For each project, we have N monthly data and apply a ``sliding-window-style'' way to build the models. The methods use training set to construct the model (using goal feature as output and all other features as input), and do the prediction on testing set.   We use previous 6 months' data to predict the current month, then make 4 consecutive predictions and return the mean of prediction errors. For example, when predicting 1 month into the future, we use data from N-6 to N-1 to predict N, use data from N-7 to N-2 to predict N-1, use data from N-8 to N-3 to predict N-2, and use data from N-9 to N-4 to predict N-3, then we return the mean of prediction errors on N, N-1, N-2 and N-3. 
  In case of untuned predictors (KNN, SVR, LNR, RFT and CART), we use all 6 previous months' data for training; For hyperparameter optimized predictors (RDCART, GSCART, FLASH and DECART), we use the first 5 of previous months' data for training, and 6th month' data for validating to find the best configuration of CART's hyperparameters (the one that achieves the closest prediction value to the actual goal on validating set), then apply this configuration on CART to make prediction on the testing set.

We run each method 20 times to reduce the bias from random operators.


\subsection{Performance Metrics}
To evaluate the performance of learners, we use two performance metrics to measure the prediction results of our experiments: Magnitude of the Relative Error (MRE) and Standardized Accuracy (SA). We use them since (a)~they are advocated in the literature~\cite{shepperd2012evaluating,sarro2016multi}; and (b)~they both offer a way to compare results against some baseline
(and such  comparisons with some baselines is considered good practice in empirical AI~\cite{Cohen95}).

Our first evaluation measure metric
is the magnitude of the relative error, or MRE. MRE is calculated by expressing absolute residual (AR) as a ratio of actual value, where AR is computed from the difference between predicted and actual values:

\[
\mathit{MRE} = \frac{|\mathit{PREDICT} - \mathit{ACTUAL}|}{\mathit{ACTUAL}}
\]

For MRE, there is the case when ACTUAL equals ``0'' and then the metric will have ``divide by zero'' error. To deal with this issue, when ACTUAL gets ``0'' in the experiment, we set MRE to ``0'' if PREDICT is also ``0'', or a value larger than ``1'' otherwise.

Sarro et al.~\cite{sarro2016multi} favors MRE
since, they argue that, it is known that the human expert performance
for certain SE estimation tasks  has a MRE of 30\% ~\cite{molokken2003review}. That is to say, if some estimators achieve less than 30\% MRE then it can be said to be competitive with human level performance.

MRE has been criticized because of its bias towards error underestimations~\cite{foss2003simulation,kitchenham2001accuracy,korte2008confidence,port2008comparative,shepperd2000building,stensrud2003further}.  
Shepperd et al. champion another evaluation measure called ``standardized accuracy'', or SA~\cite{shepperd2012evaluating}.
SA is computed as the ratio of the observed error against some reasonable fast-but-unsophisticated measurement. That is to say, 
SA expresses itself as the ratio of some sophisticated estimate
divided by a much simpler method.
SA~\cite{langdon2016exact,shepperd2012evaluating} is based on Mean Absolute Error (MAE), which is defined in terms of 

\[
\mathit{MAE}=\frac{1}{N}\sum_{i=1}^n|\mathit{PREDICT}_i-\mathit{ACTUAL}_i|
\]

where $N$ is the number of data used for evaluating the performance. SA uses MAE as follows:

\[
\mathit{SA} = (1-\frac{\mathit{MAE}}{\mathit{MAE}_{guess}})\times 100
\]

where $\mathit{MAE}_{guess}$ is the $\mathit{MAE}$ of a set of guessing values. In our case, we use the median of previous months' values as the guessing values.

We find Shepperd et al.'s arguments for SA to be compelling. But we also agree with Sarro et al. that it is useful to   compare estimates against some human-level baselines. Hence, for completeness, we apply both evaluation metrics. As shown below, both evaluation metrics will offer the same conclusion (that DECART's performance is both useful and better than other  methods for predicting project health indicators).

Note that in all our results: For MRE, {\em smaller}  values are {\em better}, and the best possible performance result is ``0''. For SA,  {\em larger} are {\em better }, the best possible performance result is ``100\%''.

\subsection{Statistics}
\label{sect:stats} 
We report the median (50th percentile) and inter-quartile range (IQR=75th-25th percentile) of our methods' performance across all 1,159 projects.  

To decide which methods do better than any other, we follow the suggestion by Dem{\v{s}}ar et al. and Herbold et al., use Friedman test with Nemenyi Post-Hoc test to differentiate the performance of each methods~\cite{demvsar2006statistical,herbold2017comments,herbold2018correction}.

Friedman test is a non-parametric statistical test which determines whether or not there is a statistically significant difference between three or more populations.~\cite{friedman1940comparison}.

Nemenyi Post-Hoc test is used to decide which groups are significantly different from each other~\cite{nemenyi1963distribution}. In case the Friedman test determines that there are
statistically significant differences between the populations, the Nemenyi test
uses Critical Distances (CD) between average ranks to define significant different populations. If the distance between two average ranks is greater than the Critical Distances (CD), these two populations will be treated as significantly different.

For each project in our experiments, each method gets a performance population when predicting an health indicator. We first run Friedman test across these populations, if the corresponding p-value is larger than $\mathit{0.05}$, we consider the difference of the methods are not significant (for this project), and we mark all methods belong to ``first group''.

If the corresponding p-value is less than $\mathit{0.05}$, we then run Nemenyi test to compare the performance populations with each method. We set the threshold of p-value in Nemenyi test to $\mathit{0.05}$ and differentiate the methods into different groups, and mark the methods in group with the best performance as ``first group''.

\section{Results}
\label{sect:resul} 
In this section, we answer the related research questions based on the experiment results.

\begin{table}[!t]
\centering
\caption{
 MRE median  results:  one month into the future.}
\label{tbl:med_mre}
\begin{adjustbox}{max width=0.96\textwidth}     
\begin{tabular}{lcccccccccc}
 & KNN & LNR & SVR & RFT & CART & RDCART & GSCART & FLASH & DECART & ASKL \\
commit & \cellcolor[HTML]{EFEFEF}51\% & \cellcolor[HTML]{F3F3F3}148\% & \cellcolor[HTML]{F1F1F1}101\% & \cellcolor[HTML]{F0F0F0}64\% & \cellcolor[HTML]{EFEFEF}54\% & \cellcolor[HTML]{EFEFEF}48\% & \cellcolor[HTML]{EFEFEF}35\% & \cellcolor[HTML]{EFEFEF}43\% & \cellcolor[HTML]{EFEFEF}39\% & \cellcolor[HTML]{EFEFEF}36\% \\
contributor & \cellcolor[HTML]{EFEFEF}38\% & \cellcolor[HTML]{EFEFEF}38\% & \cellcolor[HTML]{F0F0F0}72\% & \cellcolor[HTML]{F0F0F0}61\% & \cellcolor[HTML]{EFEFEF}38\% & \cellcolor[HTML]{EFEFEF}36\% & \cellcolor[HTML]{EFEFEF}33\% & \cellcolor[HTML]{666666}21\% & \cellcolor[HTML]{696969}21\% & \cellcolor[HTML]{838383}23\% \\
openPR & \cellcolor[HTML]{EFEFEF}40\% & \cellcolor[HTML]{EFEFEF}39\% & \cellcolor[HTML]{F0F0F0}74\% & \cellcolor[HTML]{EFEFEF}43\% & \cellcolor[HTML]{EFEFEF}32\% & \cellcolor[HTML]{CCCCCC}29\% & \cellcolor[HTML]{909090}24\% & \cellcolor[HTML]{A5A5A5}26\% & \cellcolor[HTML]{7A7A7A}22\% & \cellcolor[HTML]{767676}22\% \\
closePR & \cellcolor[HTML]{EFEFEF}55\% & \cellcolor[HTML]{F0F0F0}66\% & \cellcolor[HTML]{EFEFEF}46\% & \cellcolor[HTML]{EFEFEF}56\% & \cellcolor[HTML]{EFEFEF}52\% & \cellcolor[HTML]{EFEFEF}40\% & \cellcolor[HTML]{CBCBCB}29\% & \cellcolor[HTML]{E0E0E0}30\% & \cellcolor[HTML]{EFEFEF}33\% & \cellcolor[HTML]{EFEFEF}32\% \\
openISSUE & \cellcolor[HTML]{EFEFEF}57\% & \cellcolor[HTML]{F0F0F0}78\% & \cellcolor[HTML]{EFEFEF}56\% & \cellcolor[HTML]{EFEFEF}56\% & \cellcolor[HTML]{EFEFEF}49\% & \cellcolor[HTML]{EFEFEF}35\% & \cellcolor[HTML]{ADADAD}26\% & \cellcolor[HTML]{BEBEBE}28\% & \cellcolor[HTML]{868686}23\% & \cellcolor[HTML]{9D9D9D}25\% \\
closedISSUE & \cellcolor[HTML]{EFEFEF}60\% & \cellcolor[HTML]{F0F0F0}70\% & \cellcolor[HTML]{F0F0F0}89\% & \cellcolor[HTML]{EFEFEF}46\% & \cellcolor[HTML]{EFEFEF}50\% & \cellcolor[HTML]{EFEFEF}41\% & \cellcolor[HTML]{C4C4C4}28\% & \cellcolor[HTML]{E7E7E7}31\% & \cellcolor[HTML]{BFBFBF}28\% & \cellcolor[HTML]{9D9D9D}25\%
\end{tabular}
\end{adjustbox}   
\end{table}

\begin{table}[!t]
\centering
\caption{
MRE IQR results: one month into the future. }
\label{tbl:iqr_mre}
\begin{adjustbox}{max width=0.96\textwidth}     
\begin{tabular}{lcccccccccc}
 & KNN & LNR & SVR & RFT & CART & RDCART & GSCART & FLASH & DECART & ASKL \\
commit & \cellcolor[HTML]{F0F0F0}105\% & \cellcolor[HTML]{F2F2F2}152\% & \cellcolor[HTML]{F3F3F3}166\% & \cellcolor[HTML]{F1F1F1}122\% & \cellcolor[HTML]{F0F0F0}96\% & \cellcolor[HTML]{F0F0F0}91\% & \cellcolor[HTML]{EFEFEF}84\% & \cellcolor[HTML]{EFEFEF}75\% & \cellcolor[HTML]{EFEFEF}78\% & \cellcolor[HTML]{EFEFEF}81\% \\
contributor & \cellcolor[HTML]{EFEFEF}84\% & \cellcolor[HTML]{F0F0F0}94\% & \cellcolor[HTML]{F0F0F0}99\% & \cellcolor[HTML]{EFEFEF}67\% & \cellcolor[HTML]{EFEFEF}72\% & \cellcolor[HTML]{BCBCBC}53\% & \cellcolor[HTML]{B3B3B3}52\% & \cellcolor[HTML]{B9B9B9}53\% & \cellcolor[HTML]{8B8B8B}46\% & \cellcolor[HTML]{A6A6A6}50\% \\
openPR & \cellcolor[HTML]{EFEFEF}87\% & \cellcolor[HTML]{F0F0F0}104\% & \cellcolor[HTML]{F1F1F1}123\% & \cellcolor[HTML]{EFEFEF}77\% & \cellcolor[HTML]{EFEFEF}84\% & \cellcolor[HTML]{F0F0F0}95\% & \cellcolor[HTML]{EFEFEF}68\% & \cellcolor[HTML]{EFEFEF}83\% & \cellcolor[HTML]{EFEFEF}76\% & \cellcolor[HTML]{EFEFEF}73\% \\
closePR & \cellcolor[HTML]{F0F0F0}108\% & \cellcolor[HTML]{F1F1F1}117\% & \cellcolor[HTML]{F0F0F0}93\% & \cellcolor[HTML]{F0F0F0}92\% & \cellcolor[HTML]{F0F0F0}93\% & \cellcolor[HTML]{F0F0F0}88\% & \cellcolor[HTML]{F0F0F0}90\% & \cellcolor[HTML]{EFEFEF}85\% & \cellcolor[HTML]{EFEFEF}71\% & \cellcolor[HTML]{EFEFEF}72\% \\
openISSUE & \cellcolor[HTML]{CCCCCC}56\% & \cellcolor[HTML]{EFEFEF}84\% & \cellcolor[HTML]{F0F0F0}107\% & \cellcolor[HTML]{EFEFEF}67\% & \cellcolor[HTML]{DADADA}58\% & \cellcolor[HTML]{939393}47\% & \cellcolor[HTML]{6C6C6C}41\% & \cellcolor[HTML]{8B8B8B}46\% & \cellcolor[HTML]{7F7F7F}44\% & \cellcolor[HTML]{666666}40\% \\
closedISSUE & \cellcolor[HTML]{EFEFEF}62\% & \cellcolor[HTML]{EFEFEF}73\% & \cellcolor[HTML]{EFEFEF}69\% & \cellcolor[HTML]{CBCBCB}56\% & \cellcolor[HTML]{EDEDED}61\% & \cellcolor[HTML]{EFEFEF}61\% & \cellcolor[HTML]{ADADAD}51\% & \cellcolor[HTML]{A6A6A6}50\% & \cellcolor[HTML]{B6B6B6}52\% & \cellcolor[HTML]{7F7F7F}44\%
\end{tabular}
\end{adjustbox}   
\end{table}

\begin{table}[!t]
\centering
\caption{
 SA median  results:  one month into the future.}
\label{tbl:med_sa} 
\begin{adjustbox}{max width=0.96\textwidth}     
\begin{tabular}{lcccccccccc}
 & KNN & LNR & SVR & RFT & CART & RDCART & GSCART & FLASH & DECART & ASKL \\
commit & \cellcolor[HTML]{F1F1F1}22\% & \cellcolor[HTML]{F3F3F3}-11\% & \cellcolor[HTML]{F3F3F3}-19\% & \cellcolor[HTML]{F0F0F0}31\% & \cellcolor[HTML]{F0F0F0}36\% & \cellcolor[HTML]{F0F0F0}34\% & \cellcolor[HTML]{F0F0F0}40\% & \cellcolor[HTML]{F0F0F0}43\% & \cellcolor[HTML]{F0F0F0}41\% & \cellcolor[HTML]{EAEAEA}45\% \\
contributor & \cellcolor[HTML]{F0F0F0}38\% & \cellcolor[HTML]{F1F1F1}18\% & \cellcolor[HTML]{F1F1F1}20\% & \cellcolor[HTML]{ECECEC}45\% & \cellcolor[HTML]{F0F0F0}32\% & \cellcolor[HTML]{F0F0F0}44\% & \cellcolor[HTML]{959595}54\% & \cellcolor[HTML]{A7A7A7}52\% & \cellcolor[HTML]{666666}59\% & \cellcolor[HTML]{959595}54\% \\
openPR & \cellcolor[HTML]{F0F0F0}34\% & \cellcolor[HTML]{F1F1F1}27\% & \cellcolor[HTML]{F0F0F0}40\% & \cellcolor[HTML]{F0F0F0}44\% & \cellcolor[HTML]{CFCFCF}48\% & \cellcolor[HTML]{F0F0F0}42\% & \cellcolor[HTML]{B2B2B2}51\% & \cellcolor[HTML]{929292}54\% & \cellcolor[HTML]{A0A0A0}53\% & \cellcolor[HTML]{8C8C8C}55\% \\
closePR & \cellcolor[HTML]{F1F1F1}17\% & \cellcolor[HTML]{F3F3F3}-15\% & \cellcolor[HTML]{F2F2F2}9\% & \cellcolor[HTML]{F1F1F1}25\% & \cellcolor[HTML]{F0F0F0}31\% & \cellcolor[HTML]{F0F0F0}33\% & \cellcolor[HTML]{F0F0F0}34\% & \cellcolor[HTML]{F0F0F0}43\% & \cellcolor[HTML]{F0F0F0}39\% & \cellcolor[HTML]{F0F0F0}43\% \\
openISSUE & \cellcolor[HTML]{F1F1F1}28\% & \cellcolor[HTML]{F2F2F2}12\% & \cellcolor[HTML]{F2F2F2}3\% & \cellcolor[HTML]{F0F0F0}38\% & \cellcolor[HTML]{F0F0F0}29\% & \cellcolor[HTML]{F0F0F0}39\% & \cellcolor[HTML]{C3C3C3}49\% & \cellcolor[HTML]{F0F0F0}42\% & \cellcolor[HTML]{B1B1B1}51\% & \cellcolor[HTML]{E1E1E1}46\% \\
closedISSUE & \cellcolor[HTML]{F0F0F0}30\% & \cellcolor[HTML]{F1F1F1}13\% & \cellcolor[HTML]{F1F1F1}28\% & \cellcolor[HTML]{F0F0F0}39\% & \cellcolor[HTML]{F0F0F0}42\% & \cellcolor[HTML]{C8C8C8}49\% & \cellcolor[HTML]{C6C6C6}49\% & \cellcolor[HTML]{E1E1E1}46\% & \cellcolor[HTML]{F0F0F0}43\% & \cellcolor[HTML]{EAEAEA}45\%
\end{tabular}
\end{adjustbox}
\end{table}

\begin{table}[!t]
\centering
\caption{
SA  IQR  results: one month into the future.}
\label{tbl:iqr_sa}
\begin{adjustbox}{max width=0.96\textwidth}          
\begin{tabular}{lcccccccccc}
 & KNN & LNR & SVR & RFT & CART & RDCART & GSCART & FLASH & DECART & ASKL \\
commit & \cellcolor[HTML]{F0F0F0}160\% & \cellcolor[HTML]{F1F1F1}206\% & \cellcolor[HTML]{F3F3F3}296\% & \cellcolor[HTML]{EFEFEF}140\% & \cellcolor[HTML]{EFEFEF}115\% & \cellcolor[HTML]{EFEFEF}104\% & \cellcolor[HTML]{9F9F9F}91\% & \cellcolor[HTML]{EFEFEF}115\% & \cellcolor[HTML]{EFEFEF}98\% & \cellcolor[HTML]{EFEFEF}101\% \\
contributor & \cellcolor[HTML]{EFEFEF}102\% & \cellcolor[HTML]{F0F0F0}174\% & \cellcolor[HTML]{F0F0F0}155\% & \cellcolor[HTML]{EFEFEF}111\% & \cellcolor[HTML]{EFEFEF}106\% & \cellcolor[HTML]{EFEFEF}98\% & \cellcolor[HTML]{C1C1C1}94\% & \cellcolor[HTML]{EFEFEF}98\% & \cellcolor[HTML]{888888}89\% & \cellcolor[HTML]{939393}90\% \\
openPR & \cellcolor[HTML]{F0F0F0}151\% & \cellcolor[HTML]{EFEFEF}108\% & \cellcolor[HTML]{EFEFEF}117\% & \cellcolor[HTML]{EFEFEF}107\% & \cellcolor[HTML]{EFEFEF}124\% & \cellcolor[HTML]{EFEFEF}110\% & \cellcolor[HTML]{EFEFEF}98\% & \cellcolor[HTML]{EFEFEF}105\% & \cellcolor[HTML]{9B9B9B}91\% & \cellcolor[HTML]{888888}89\% \\
closePR & \cellcolor[HTML]{EFEFEF}100\% & \cellcolor[HTML]{F0F0F0}179\% & \cellcolor[HTML]{F0F0F0}184\% & \cellcolor[HTML]{EFEFEF}105\% & \cellcolor[HTML]{EFEFEF}120\% & \cellcolor[HTML]{EFEFEF}104\% & \cellcolor[HTML]{EFEFEF}103\% & \cellcolor[HTML]{C1C1C1}94\% & \cellcolor[HTML]{EFEFEF}102\% & \cellcolor[HTML]{EFEFEF}107\% \\
openISSUE & \cellcolor[HTML]{F0F0F0}150\% & \cellcolor[HTML]{F1F1F1}222\% & \cellcolor[HTML]{F2F2F2}291\% & \cellcolor[HTML]{F0F0F0}163\% & \cellcolor[HTML]{EFEFEF}136\% & \cellcolor[HTML]{E3E3E3}97\% & \cellcolor[HTML]{B4B4B4}93\% & \cellcolor[HTML]{E3E3E3}97\% & \cellcolor[HTML]{D8D8D8}96\% & \cellcolor[HTML]{E3E3E3}97\% \\
closedISSUE & \cellcolor[HTML]{EFEFEF}147\% & \cellcolor[HTML]{F0F0F0}160\% & \cellcolor[HTML]{F0F0F0}186\% & \cellcolor[HTML]{EFEFEF}142\% & \cellcolor[HTML]{EFEFEF}143\% & \cellcolor[HTML]{EFEFEF}116\% & \cellcolor[HTML]{A5A5A5}92\% & \cellcolor[HTML]{666666}86\% & \cellcolor[HTML]{888888}89\% & \cellcolor[HTML]{C1C1C1}94\%
\end{tabular}
\end{adjustbox}
\end{table}

\subsection{How to find the trends of project health indicators? (RQ1)}

We predict the value of health indicators for recent months using data from previous months.
The median and IQR values of performance results in terms of MRE and SA are shown in Table~\ref{tbl:med_mre}, Table~\ref{tbl:iqr_mre},  Table~\ref{tbl:med_sa}, and Table~\ref{tbl:iqr_sa}, respectively.

In all these four tables, we show median  and IQR of performance results across 1,159 projects.
For MRE, {\em lower} values are {\em better}, the dark cells denote better results; For SA, {\em higher} values are {\em better}, and dark cells denote better results.

\begin{table}[!t]
\centering
\caption{MRE and SA results with DECART, predicting for 1, 3, 6, 12 months
into the future.}
\label{tbl:change}
\begin{adjustbox}{max width=0.96\textwidth}     
\begin{tabular}{llrrrr}
 & Health Indicator & \multicolumn{1}{l}{1 month} & \multicolumn{1}{l}{3 month} & \multicolumn{1}{l}{6 month} & \multicolumn{1}{l}{12 month} \\ \hline
 & commit & 0.39 & 0.40 & 0.43 & 0.53 \\
 & contributor & 0.21 & 0.21 & 0.23 & 0.26 \\
 & openPR & 0.22 & 0.22 & 0.24 & 0.30 \\
 & closePR & 0.33 & 0.33 & 0.35 & 0.40 \\
 & openISSUE & 0.23 & 0.23 & 0.26 & 0.30 \\
 & closedISSUE & 0.28 & 0.29 & 0.30 & 0.35 \\
\multirow{-7}{*}{Median, MRE} & \cellcolor[HTML]{CCCCCC}median ratio change & \multicolumn{1}{l}{\cellcolor[HTML]{CCCCCC}} & \cellcolor[HTML]{CCCCCC}101\% & \cellcolor[HTML]{CCCCCC}107\% & \cellcolor[HTML]{CCCCCC}117\% \\ \hline
 & commit & 0.41 & \cellcolor[HTML]{FFFFFF}0.40 & \cellcolor[HTML]{FFFFFF}0.38 & \cellcolor[HTML]{FFFFFF}0.30 \\
 & contributor & 0.59 & \cellcolor[HTML]{FFFFFF}0.57 & \cellcolor[HTML]{FFFFFF}0.51 & \cellcolor[HTML]{FFFFFF}0.40 \\
 & openPR & 0.53 & \cellcolor[HTML]{FFFFFF}0.52 & \cellcolor[HTML]{FFFFFF}0.48 & \cellcolor[HTML]{FFFFFF}0.38 \\
 & closePR & 0.39 & \cellcolor[HTML]{FFFFFF}0.38 & \cellcolor[HTML]{FFFFFF}0.34 & \cellcolor[HTML]{FFFFFF}0.28 \\
 & openISSUE & 0.51 & \cellcolor[HTML]{FFFFFF}0.50 & \cellcolor[HTML]{FFFFFF}0.44 & \cellcolor[HTML]{FFFFFF}0.36 \\
 & closedISSUE & 0.43 & \cellcolor[HTML]{FFFFFF}0.43 & \cellcolor[HTML]{FFFFFF}0.39 & \cellcolor[HTML]{FFFFFF}0.32 \\
\multirow{-7}{*}{Median, SA} & \cellcolor[HTML]{CCCCCC}median ratio change & \multicolumn{1}{l}{\cellcolor[HTML]{CCCCCC}} & \cellcolor[HTML]{CCCCCC}98\% & \cellcolor[HTML]{CCCCCC}91\% & \cellcolor[HTML]{CCCCCC}80\% \\ \hline
 & commit & 0.78 & \cellcolor[HTML]{FFFFFF}0.85 & \cellcolor[HTML]{FFFFFF}0.99 & \cellcolor[HTML]{FFFFFF}1.11 \\
 & contributor & 0.46 & \cellcolor[HTML]{FFFFFF}0.49 & \cellcolor[HTML]{FFFFFF}0.57 & \cellcolor[HTML]{FFFFFF}0.66 \\
 & openPR & 0.76 & \cellcolor[HTML]{FFFFFF}0.82 & \cellcolor[HTML]{FFFFFF}0.94 & \cellcolor[HTML]{FFFFFF}1.07 \\
 & closePR & 0.71 & \cellcolor[HTML]{FFFFFF}0.76 & \cellcolor[HTML]{FFFFFF}0.90 & \cellcolor[HTML]{FFFFFF}1.02 \\
 & openISSUE & 0.44 & \cellcolor[HTML]{FFFFFF}0.47 & \cellcolor[HTML]{FFFFFF}0.54 & \cellcolor[HTML]{FFFFFF}0.61 \\
 & closedISSUE & 0.52 & \cellcolor[HTML]{FFFFFF}0.56 & \cellcolor[HTML]{FFFFFF}0.66 & \cellcolor[HTML]{FFFFFF}0.75 \\
\multirow{-7}{*}{IQR, MRE} & \cellcolor[HTML]{CCCCCC}median ratio change & \multicolumn{1}{l}{\cellcolor[HTML]{CCCCCC}} & \cellcolor[HTML]{CCCCCC}108\% & \cellcolor[HTML]{CCCCCC}116\% & \cellcolor[HTML]{CCCCCC}113\% \\ \hline
 & commit & 0.98 & \cellcolor[HTML]{FFFFFF}1.07 & \cellcolor[HTML]{FFFFFF}1.19 & \cellcolor[HTML]{FFFFFF}1.47 \\
 & contributor & 0.89 & \cellcolor[HTML]{FFFFFF}0.97 & \cellcolor[HTML]{FFFFFF}1.10 & \cellcolor[HTML]{FFFFFF}1.35 \\
 & openPR & 0.91 & \cellcolor[HTML]{FFFFFF}0.99 & \cellcolor[HTML]{FFFFFF}1.12 & \cellcolor[HTML]{FFFFFF}1.35 \\
 & closePR & 1.02 & \cellcolor[HTML]{FFFFFF}1.12 & \cellcolor[HTML]{FFFFFF}1.28 & \cellcolor[HTML]{FFFFFF}1.57 \\
 & openISSUE & 0.96 & \cellcolor[HTML]{FFFFFF}1.04 & \cellcolor[HTML]{FFFFFF}1.14 & \cellcolor[HTML]{FFFFFF}1.34 \\
 & closedISSUE & 0.89 & \cellcolor[HTML]{FFFFFF}0.97 & \cellcolor[HTML]{FFFFFF}1.08 & \cellcolor[HTML]{FFFFFF}1.25 \\
\multirow{-7}{*}{IQR, SA} & \cellcolor[HTML]{CCCCCC}median ratio change & \multicolumn{1}{l}{\cellcolor[HTML]{CCCCCC}} & \cellcolor[HTML]{CCCCCC}109\% & \cellcolor[HTML]{CCCCCC}112\% & \cellcolor[HTML]{CCCCCC}122\% \\ \hline
\end{tabular}
\end{adjustbox}
\end{table}

In these results, we observe that our methods provide very different performance with these 6 health indicators' prediction. In Table~\ref{tbl:med_mre}, we see that some learners have errors over 100\% (LNR, predicting for number of commits). For the same task, other learners, however, only have around half of the errors (CART, 54\%).
Also in that table, when predicting number of commits, the median MRE scores of the untuned learners (KNN, LNR, SVR, RFT, CART) are over 50\%. That is,  these estimates are often wrong by a factor of two, or more.
Further, these tables show that   hyperparameter optimization is   beneficial.
The  GSCART, FLASH, DECART and ASKL columns of Table~\ref{tbl:med_mre} and Table~\ref{tbl:med_sa} show that these methods have better median MREs and SAs than the untuned methods. For example, as shown in Table~\ref{tbl:med_mre}, the median error for DECART is usually at least 15\% less than the CART. Additionally, the results of Table~\ref{tbl:iqr_mre} and Table~\ref{tbl:iqr_sa} also demonstrate the stability of HPO methods (with the lower IQR when measuring the performance variability of all methods).


Turning now to other prediction results, our next set of results shows what happens when we make predictions over a 1, 3, 6, 12 months interval.
Note that to simulate predicting the status of ahead $1st$, $3rd$, $6th$, $12th$ month, for a project with $N$ months of data, the training sets need to be selected from month 1 to month $N-1$, $N-3$, $N-6$, $N-12$, respectively. That is, to say that the {\em further} ahead of our predictions, the {\em earlier} data we have for training. Hence, one thing to watch for is whether or not performance decreases as the training set ages.


  Table~\ref{tbl:change} presents the MRE and SA results of DECART, predicting for 1, 3, 6, and 12 months into the future.
  By observing the median of ratio-changing (show in gray) from left to right across the table, we see that as we try to predict further and further into the future, (a)~MRE degrades around 17\% and (b)~SA degrades only about 20\%, or less.
 Measured in absolute terms, these changes are still relatively small. 
In any case, summarizing all the above, we say that:

\begin{blockquote}
\noindent
\textbf{Answer 1}: Many project health indicators can be predicted, with good accuracy, for 1, 3, 6, 12   months into the future.
For example, we can predict the number of contributors next month with an error of 21\% (MRE).
\end{blockquote}

\begin{figure}[!b]
\begin{center}
\includegraphics[width=4.5in]{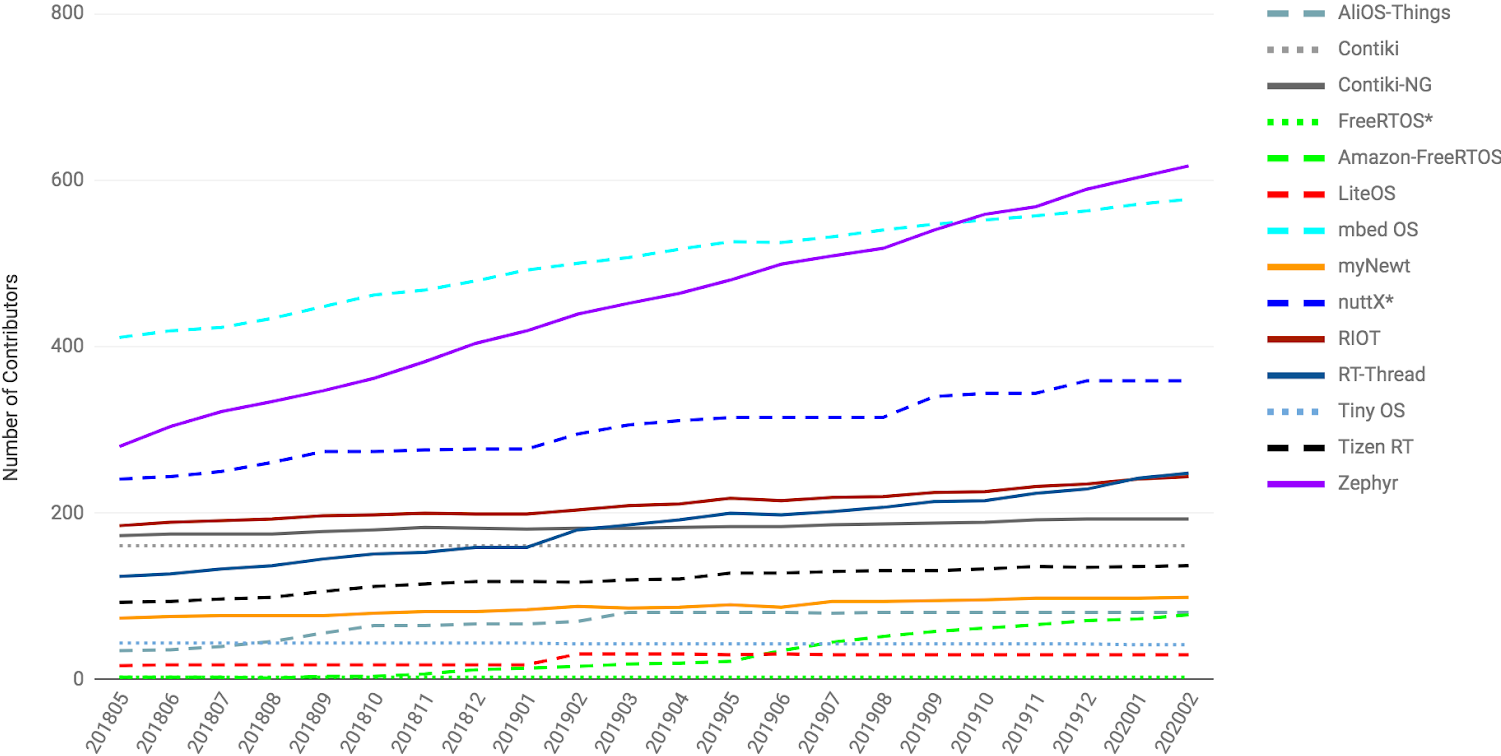}
\end{center}
 \caption{The trends in number of contributors across 14 real-time OS projects since 2018. Zephyr (one in purple) has 
  out-paced other similar software projects. (data source: The Apache Software Foundation)}\label{fig:apache}
 \end{figure}

\begin{table*}[!t]
\centering

\begin{adjustbox}{max width=0.96\textwidth}
\begin{tabular}{r|cccccccccc}
\rowcolor[HTML]{FFFFFF}
{\color[HTML]{000000} Target } & \multicolumn{1}{l}{{\color[HTML]{000000} \begin{sideways}commit\end{sideways}}} & \multicolumn{1}{l}{{\color[HTML]{000000} \begin{sideways}contributor\end{sideways}}} & \multicolumn{1}{l}{{\color[HTML]{000000} \begin{sideways}openPR\end{sideways}}} & \multicolumn{1}{l}{{\color[HTML]{000000} \begin{sideways}closePR\end{sideways}}} & \multicolumn{1}{l}{{\color[HTML]{000000} \begin{sideways}openISSUE\end{sideways}}} & \multicolumn{1}{l}{{\color[HTML]{000000} \begin{sideways}closeISSUE\end{sideways}}} & \multicolumn{1}{l}{{\color[HTML]{000000} \begin{sideways}star\end{sideways}}} & \multicolumn{1}{l}{{\color[HTML]{000000} \begin{sideways}ISSUEcomment\end{sideways}}} & \multicolumn{1}{l}{{\color[HTML]{000000} \begin{sideways}mergedPR\end{sideways}}} & \multicolumn{1}{l}{{\color[HTML]{000000} \begin{sideways}fork\end{sideways}}} \\ \hline
commit & \cellcolor[HTML]{F4CCCC}n/a & \cellcolor[HTML]{D9D9D9}7\% & \cellcolor[HTML]{9C9C9C}15\% & \cellcolor[HTML]{F3F3F3}4\% & \cellcolor[HTML]{D9D9D9}7\% & \cellcolor[HTML]{666666}22\% & \cellcolor[HTML]{E2E2E2}6\% & \cellcolor[HTML]{7D7D7D}19\% & \cellcolor[HTML]{EBEBEB}5\% & \cellcolor[HTML]{9C9C9C}15\% \\
contributor & \cellcolor[HTML]{6E6E6E}22\% & \cellcolor[HTML]{F4CCCC}n/a & \cellcolor[HTML]{F3F3F3}4\% & \cellcolor[HTML]{F3F3F3}4\% & \cellcolor[HTML]{7E7E7E}20\% & \cellcolor[HTML]{D9D9D9}8\% & \cellcolor[HTML]{666666}23\% & \cellcolor[HTML]{EDEDED}5\% & \cellcolor[HTML]{D9D9D9}8\% & \cellcolor[HTML]{E7E7E7}6\% \\
openPR & \cellcolor[HTML]{CACACA}10\% & \cellcolor[HTML]{D9D9D9}7\% & \cellcolor[HTML]{F4CCCC}n/a & \cellcolor[HTML]{E4E4E4}5\% & \cellcolor[HTML]{818181}24\% & \cellcolor[HTML]{666666}29\% & \cellcolor[HTML]{F3F3F3}2\% & \cellcolor[HTML]{E9E9E9}4\% & \cellcolor[HTML]{AAAAAA}16\% & \cellcolor[HTML]{EEEEEE}3\% \\
closePR & \cellcolor[HTML]{BBBBBB}13\% & \cellcolor[HTML]{C3C3C3}12\% & \cellcolor[HTML]{666666}24\% & \cellcolor[HTML]{F4CCCC}n/a & \cellcolor[HTML]{D9D9D9}9\% & \cellcolor[HTML]{7D7D7D}21\% & \cellcolor[HTML]{E4E4E4}7\% & \cellcolor[HTML]{E9E9E9}6\% & \cellcolor[HTML]{F3F3F3}4\% & \cellcolor[HTML]{F3F3F3}4\% \\
openISSUE & \cellcolor[HTML]{EDEDED}3\% & \cellcolor[HTML]{D9D9D9}6\% & \cellcolor[HTML]{9A9A9A}16\% & \cellcolor[HTML]{F3F3F3}2\% & \cellcolor[HTML]{F4CCCC}n/a & \cellcolor[HTML]{9A9A9A}16\% & \cellcolor[HTML]{E0E0E0}5\% & \cellcolor[HTML]{E0E0E0}5\% & \cellcolor[HTML]{666666}24\% & \cellcolor[HTML]{6D6D6D}23\% \\
closedISSUE & \cellcolor[HTML]{D2D2D2}9\% & \cellcolor[HTML]{E2E2E2}7\% & \cellcolor[HTML]{666666}24\% & \cellcolor[HTML]{B6B6B6}13\% & \cellcolor[HTML]{D9D9D9}8\% & \cellcolor[HTML]{F4CCCC}n/a & \cellcolor[HTML]{F3F3F3}5\% & \cellcolor[HTML]{7C7C7C}21\% & \cellcolor[HTML]{EBEBEB}6\% & \cellcolor[HTML]{E2E2E2}7\% \\ \hline
\rowcolor[HTML]{FFFFFF} 
mean & 11\% & 8\% & 17\% & 6\% & 14\% & 19\% & 8\% & 10\% & 11\% & 10\%
\end{tabular}
\end{adjustbox}
\caption{
The mean scores of Gini importance in trees generated by DECART (observed percentages in 1,159 cases).}
\label{tbl:frq}
\end{table*}

To give a scenario where this kind of prediction would be useful, we provide a real-world use case with a set of 14 real time OS projects currently supported by the Apache Software Foundation. As shown in Figure~\ref{fig:apache},  one of these operating systems, Zephyr (the one in purple), is exhibiting the steepest growth in the number of its contributors. As to the other projects, most
of these might be viewed as stagnant, perhaps even at risk of cancellation\footnote{In the Apache Software Foundation, projects can be canceled and  ``moved to the attic'' (https://attic.apache.org) when they are unable to muster 3 votes for a release, lack of active contributors, or unable to fulfill their reporting duties to the Foundation.}. 
Hence, to the managers of the stagnant
projects in Figure~\ref{fig:apache} (those with mostly flat curves),  they are particularly interested in ``catching up with  Zephyr''. Note that this
means increasing
their number of contributors to the ``flat`` projects by 200\% (for RIOT) to 3,000\% (LiteOS). For that purpose,
predictions with a mere 21\% error (next month) would be useful
to foretell improvements to the software.

\subsection{What features matter the most in prediction? (RQ2)}
In our experimental data, we have 10 numeric features for prediction.
We use them since they are features with high importance, suggested by prior work (see Section~\ref{sect:data_collect}).
That said, having done all these experiments, it makes sense to ask which features, in practice, would be more useful when we predict health indicators.   This information could help us to focus on useful features and remove irrelevancies when enlarging our research in future work. To work that out, we look into the trees generated by DECART
(one of our best learners) in the above experiments. 
For each tree, we find impurity-based feature importances, which is computed as the normalized total reduction of the criterion brought by the feature (also known as the Gini importance).

For each predicting target, we calculate the mean scores for each feature, of all generated trees, the results are summarized in  Table~\ref{tbl:frq}. 
In this table, ``n/a''  denotes the dependent variable which is not counted in the experiment. From this table, first of all, we find that some features are highly related to specific health indicators. For example, ``commit'', ``openISSUE'' and  ``star'' have got scores  $22\%$,  $20\%$ and  $23\%$  when we built trees to predict ``contributor'' indicator for $1,159$ repositories.  Secondly, some features are bellwethers that have been used as features for multiple indicator predictions, like ``openPR'' gets $15\%$, $24\%$, $16\%$, and $24\%$ when predicting ``commit'', ``closePR'', ``openISSUE'' and ``closeISSUE''. Thirdly, some features even though they belong to the similar type, like ``openISSUE'' and ``closeISSUE'', they are not highly related in the predictions. In our experiment, we find that  ``openISSUE'' only gets $8\%$, way less than ``ISSUEcomment'' ($21\%$), ``openPR'' ($24\%$) and ``closePR'' ($13\%$) when predicting ``closeISSUE''. Last but not least,  some features are less used than others. According to our experiment, ``closePR'' is the least used feature for all predictions (the mean score of ``closePR'' is only $6\%$).

\begin{blockquote}
\noindent
\textbf{Answer 2}: In our study, ``monthly\_commits'', ``monthly\_openPR'', ``monthly\_openISSUE'' and ``monthly\_closeISSUE'' are the most important features, while ``monthly\_closePR'' is the least used feature for all six health indicators' predictions.
\end{blockquote}

Note that none of these features should be abandoned. For feature ``closePR'', the least used feature in prediction, when predicting ``closeISSUE'', this feature still gets 13\% score of these cases.

 That said, it would be hard pressed to say
 that Table~\ref{tbl:frq} indicates that only a small subset of the \tbl{feature} features are outstandingly most important. While Table~\ref{tbl:frq} suggests that some feature pruning might be useful, overall we would suggest that using all of these features might be the best practice in most cases.

\subsection{How to improve the performance of health indicators prediction? (RQ3)}
To answer this question, 
we compared the experimental results of each method on all 1,159 open-source projects predicting for 1, 3, 6, and 12 months into the future.

Across 1,159 projects, for each health indicator and each of future month (1, 3, 6, 12), we report the ``win rate'', which are the percentages of one learner belongs to the group with the best prediction performance (i.e. in first group). As introduced in Section~\ref{sect:stats}, when predicting one health indicator, for each project, learners get their own performance populations, we use the Friedman test with Nemenyi Post-Hoc test to compare different learners' performance populations in terms of MRE and SA, then differentiate the learners into different groups, group with lower MRE (or larger SA) is the first group. For example, when predicting ``number of commits'' in ``1 month into the future'', in terms of MRE, method ``KNN'' gets into first group in 510 out of 1159 cases, while method ``DECART'' gets into first group in 939 out of 1159 cases. Hence, their win rates for predicting ``number of commits'' in ``1 month into the future'' in terms of MRE are 44\% and 81\%, respectively.

One note is, there may be multiple methods belong to the first group. \tbl{win_mre} and \tbl{win_sa} show the results of learners' win rate in terms of MRE and SA.

The comparisons in these tables are for intra-row results, where the darker cells indicate the learning methods with higher win rate. For example, in the first row of \tbl{win_mre} (except the header row), when predicting the number of commits in next month, KNN has the best MRE performance in 44\% of all 1,159 cases, and DECART has the best MRE performance in 81\% of all 1,159 cases. 

As shown in \tbl{win_mre}, in terms of MRE, ASKL and DECART achieves the best performance with winning rates from $52\%$ to $97\%$ for all predictions (the median win rate is 75\%). 
Meanwhile, the winning rates of other learners, mostly range from $20\%$ to $60\%$. For example, FLASH, the hyperparameter-optimized method used in the previous effort estimation study, no longer works the best and its median win rate is only 60\%. 

\begin{table}[!t]
\centering
\caption{Statistical analysis of the MRE results: the win rate (ranked first using the Friedman Nemenyi test of Section~\ref{sect:stats} for the    different treatments).
Raw performance measured in terms of MRE for predictions $N$ months into the future.}
\label{tbl:win_mre}
\resizebox{0.98\textwidth}{!}{
\begin{tabular}{llcccccccccc}
\multicolumn{1}{c}{Months} & \multicolumn{1}{c}{Health Indicator} & KNN & LNR & SVR & RFT & CART & RDCART & GSCART & FLASH & DECART & ASKL \\ \hline
 & commit & \cellcolor[HTML]{CFCFCF}44\% & \cellcolor[HTML]{F1F1F1}17\% & \cellcolor[HTML]{E3E3E3}28\% & \cellcolor[HTML]{D2D2D2}41\% & \cellcolor[HTML]{D9D9D9}36\% & \cellcolor[HTML]{CDCDCD}46\% & \cellcolor[HTML]{8B8B8B}79\% & \cellcolor[HTML]{A0A0A0}68\% & \cellcolor[HTML]{868686}81\% & \cellcolor[HTML]{838383}83\% \\
 & contributor & \cellcolor[HTML]{BEBEBE}53\% & \cellcolor[HTML]{DDDDDD}33\% & \cellcolor[HTML]{CDCDCD}46\% & \cellcolor[HTML]{CFCFCF}44\% & \cellcolor[HTML]{D4D4D4}40\% & \cellcolor[HTML]{CECECE}45\% & \cellcolor[HTML]{9D9D9D}70\% & \cellcolor[HTML]{C3C3C3}51\% & \cellcolor[HTML]{A1A1A1}68\% & \cellcolor[HTML]{959595}74\% \\
 & openPR & \cellcolor[HTML]{D2D2D2}41\% & \cellcolor[HTML]{CFCFCF}44\% & \cellcolor[HTML]{DEDEDE}32\% & \cellcolor[HTML]{C2C2C2}51\% & \cellcolor[HTML]{C4C4C4}50\% & \cellcolor[HTML]{ACACAC}62\% & \cellcolor[HTML]{9F9F9F}69\% & \cellcolor[HTML]{B7B7B7}57\% & \cellcolor[HTML]{747474}90\% & \cellcolor[HTML]{767676}89\% \\
 & closePR & \cellcolor[HTML]{D7D7D7}38\% & \cellcolor[HTML]{E4E4E4}27\% & \cellcolor[HTML]{BEBEBE}53\% & \cellcolor[HTML]{CCCCCC}46\% & \cellcolor[HTML]{C8C8C8}48\% & \cellcolor[HTML]{CBCBCB}47\% & \cellcolor[HTML]{C3C3C3}51\% & \cellcolor[HTML]{B2B2B2}59\% & \cellcolor[HTML]{BDBDBD}54\% & \cellcolor[HTML]{B1B1B1}60\% \\
 & openISSUE & \cellcolor[HTML]{C1C1C1}52\% & \cellcolor[HTML]{E0E0E0}31\% & \cellcolor[HTML]{EAEAEA}22\% & \cellcolor[HTML]{D8D8D8}37\% & \cellcolor[HTML]{CFCFCF}44\% & \cellcolor[HTML]{C5C5C5}50\% & \cellcolor[HTML]{878787}81\% & \cellcolor[HTML]{AFAFAF}61\% & \cellcolor[HTML]{9E9E9E}69\% & \cellcolor[HTML]{989898}72\% \\
\multirow{-6}{*}{1st} & closedISSUE & \cellcolor[HTML]{D6D6D6}39\% & \cellcolor[HTML]{E3E3E3}28\% & \cellcolor[HTML]{F0F0F0}17\% & \cellcolor[HTML]{D7D7D7}37\% & \cellcolor[HTML]{C9C9C9}48\% & \cellcolor[HTML]{D2D2D2}42\% & \cellcolor[HTML]{B2B2B2}59\% & \cellcolor[HTML]{C4C4C4}50\% & \cellcolor[HTML]{B5B5B5}58\% & \cellcolor[HTML]{ABABAB}63\% \\ \hline
 & commit & \cellcolor[HTML]{DDDDDD}33\% & \cellcolor[HTML]{E2E2E2}29\% & \cellcolor[HTML]{E0E0E0}30\% & \cellcolor[HTML]{CACACA}47\% & \cellcolor[HTML]{D0D0D0}43\% & \cellcolor[HTML]{C4C4C4}50\% & \cellcolor[HTML]{B3B3B3}59\% & \cellcolor[HTML]{AFAFAF}61\% & \cellcolor[HTML]{838383}83\% & \cellcolor[HTML]{8C8C8C}78\% \\
 & contributor & \cellcolor[HTML]{BDBDBD}54\% & \cellcolor[HTML]{E4E4E4}27\% & \cellcolor[HTML]{ECECEC}21\% & \cellcolor[HTML]{D7D7D7}38\% & \cellcolor[HTML]{DEDEDE}32\% & \cellcolor[HTML]{DADADA}35\% & \cellcolor[HTML]{898989}80\% & \cellcolor[HTML]{BDBDBD}54\% & \cellcolor[HTML]{727272}91\% & \cellcolor[HTML]{717171}92\% \\
 & openPR & \cellcolor[HTML]{E8E8E8}24\% & \cellcolor[HTML]{D2D2D2}41\% & \cellcolor[HTML]{DFDFDF}31\% & \cellcolor[HTML]{D0D0D0}43\% & \cellcolor[HTML]{D5D5D5}39\% & \cellcolor[HTML]{C8C8C8}48\% & \cellcolor[HTML]{BFBFBF}53\% & \cellcolor[HTML]{AFAFAF}61\% & \cellcolor[HTML]{A4A4A4}66\% & \cellcolor[HTML]{9F9F9F}69\% \\
 & closePR & \cellcolor[HTML]{D8D8D8}37\% & \cellcolor[HTML]{E7E7E7}24\% & \cellcolor[HTML]{D0D0D0}43\% & \cellcolor[HTML]{D5D5D5}39\% & \cellcolor[HTML]{BCBCBC}54\% & \cellcolor[HTML]{BDBDBD}54\% & \cellcolor[HTML]{898989}80\% & \cellcolor[HTML]{A4A4A4}66\% & \cellcolor[HTML]{A3A3A3}67\% & \cellcolor[HTML]{8C8C8C}78\% \\
 & openISSUE & \cellcolor[HTML]{CECECE}45\% & \cellcolor[HTML]{EDEDED}20\% & \cellcolor[HTML]{E3E3E3}28\% & \cellcolor[HTML]{D2D2D2}41\% & \cellcolor[HTML]{D5D5D5}39\% & \cellcolor[HTML]{AFAFAF}61\% & \cellcolor[HTML]{B0B0B0}60\% & \cellcolor[HTML]{C4C4C4}50\% & \cellcolor[HTML]{AFAFAF}61\% & \cellcolor[HTML]{A6A6A6}65\% \\
\multirow{-6}{*}{3rd} & closedISSUE & \cellcolor[HTML]{DBDBDB}34\% & \cellcolor[HTML]{E8E8E8}24\% & \cellcolor[HTML]{E9E9E9}23\% & \cellcolor[HTML]{D0D0D0}43\% & \cellcolor[HTML]{D7D7D7}38\% & \cellcolor[HTML]{D4D4D4}40\% & \cellcolor[HTML]{6D6D6D}94\% & \cellcolor[HTML]{9C9C9C}70\% & \cellcolor[HTML]{818181}84\% & \cellcolor[HTML]{909090}76\% \\ \hline
 & commit & \cellcolor[HTML]{D5D5D5}39\% & \cellcolor[HTML]{E4E4E4}27\% & \cellcolor[HTML]{E1E1E1}29\% & \cellcolor[HTML]{D9D9D9}36\% & \cellcolor[HTML]{C8C8C8}48\% & \cellcolor[HTML]{B4B4B4}58\% & \cellcolor[HTML]{959595}74\% & \cellcolor[HTML]{C8C8C8}48\% & \cellcolor[HTML]{757575}90\% & \cellcolor[HTML]{808080}84\% \\
 & contributor & \cellcolor[HTML]{CDCDCD}46\% & \cellcolor[HTML]{D9D9D9}36\% & \cellcolor[HTML]{EDEDED}20\% & \cellcolor[HTML]{DEDEDE}32\% & \cellcolor[HTML]{BBBBBB}55\% & \cellcolor[HTML]{BBBBBB}55\% & \cellcolor[HTML]{B5B5B5}58\% & \cellcolor[HTML]{BCBCBC}54\% & \cellcolor[HTML]{ADADAD}62\% & \cellcolor[HTML]{A6A6A6}65\% \\
 & openPR & \cellcolor[HTML]{D9D9D9}36\% & \cellcolor[HTML]{DEDEDE}32\% & \cellcolor[HTML]{DDDDDD}33\% & \cellcolor[HTML]{D7D7D7}38\% & \cellcolor[HTML]{CECECE}45\% & \cellcolor[HTML]{CACACA}47\% & \cellcolor[HTML]{7F7F7F}85\% & \cellcolor[HTML]{AEAEAE}61\% & \cellcolor[HTML]{838383}83\% & \cellcolor[HTML]{808080}84\% \\
 & closePR & \cellcolor[HTML]{E6E6E6}26\% & \cellcolor[HTML]{DDDDDD}32\% & \cellcolor[HTML]{DDDDDD}33\% & \cellcolor[HTML]{D1D1D1}42\% & \cellcolor[HTML]{CACACA}47\% & \cellcolor[HTML]{BFBFBF}53\% & \cellcolor[HTML]{929292}75\% & \cellcolor[HTML]{AEAEAE}61\% & \cellcolor[HTML]{BCBCBC}54\% & \cellcolor[HTML]{9D9D9D}70\% \\
 & openISSUE & \cellcolor[HTML]{D3D3D3}41\% & \cellcolor[HTML]{F1F1F1}17\% & \cellcolor[HTML]{E3E3E3}28\% & \cellcolor[HTML]{D2D2D2}42\% & \cellcolor[HTML]{D0D0D0}43\% & \cellcolor[HTML]{B5B5B5}58\% & \cellcolor[HTML]{B3B3B3}59\% & \cellcolor[HTML]{BABABA}55\% & \cellcolor[HTML]{8D8D8D}78\% & \cellcolor[HTML]{8E8E8E}77\% \\
\multirow{-6}{*}{6th} & closedISSUE & \cellcolor[HTML]{C8C8C8}48\% & \cellcolor[HTML]{E3E3E3}28\% & \cellcolor[HTML]{E7E7E7}25\% & \cellcolor[HTML]{D4D4D4}40\% & \cellcolor[HTML]{C9C9C9}48\% & \cellcolor[HTML]{C6C6C6}49\% & \cellcolor[HTML]{C1C1C1}52\% & \cellcolor[HTML]{CBCBCB}47\% & \cellcolor[HTML]{B8B8B8}56\% & \cellcolor[HTML]{C0C0C0}52\% \\ \hline
 & commit & \cellcolor[HTML]{D0D0D0}43\% & \cellcolor[HTML]{F1F1F1}17\% & \cellcolor[HTML]{E6E6E6}25\% & \cellcolor[HTML]{CDCDCD}46\% & \cellcolor[HTML]{DBDBDB}34\% & \cellcolor[HTML]{B8B8B8}56\% & \cellcolor[HTML]{A3A3A3}67\% & \cellcolor[HTML]{C2C2C2}51\% & \cellcolor[HTML]{979797}73\% & \cellcolor[HTML]{8B8B8B}79\% \\
 & contributor & \cellcolor[HTML]{C4C4C4}50\% & \cellcolor[HTML]{EAEAEA}22\% & \cellcolor[HTML]{E1E1E1}30\% & \cellcolor[HTML]{D7D7D7}37\% & \cellcolor[HTML]{D6D6D6}38\% & \cellcolor[HTML]{B6B6B6}57\% & \cellcolor[HTML]{B4B4B4}58\% & \cellcolor[HTML]{BFBFBF}53\% & \cellcolor[HTML]{BEBEBE}53\% & \cellcolor[HTML]{A2A2A2}67\% \\
 & openPR & \cellcolor[HTML]{D3D3D3}40\% & \cellcolor[HTML]{E3E3E3}28\% & \cellcolor[HTML]{E3E3E3}28\% & \cellcolor[HTML]{B4B4B4}58\% & \cellcolor[HTML]{BDBDBD}54\% & \cellcolor[HTML]{B8B8B8}56\% & \cellcolor[HTML]{868686}81\% & \cellcolor[HTML]{A9A9A9}64\% & \cellcolor[HTML]{666666}97\% & \cellcolor[HTML]{9B9B9B}71\% \\
 & closePR & \cellcolor[HTML]{C5C5C5}50\% & \cellcolor[HTML]{E6E6E6}26\% & \cellcolor[HTML]{D4D4D4}40\% & \cellcolor[HTML]{D4D4D4}40\% & \cellcolor[HTML]{D6D6D6}39\% & \cellcolor[HTML]{C6C6C6}49\% & \cellcolor[HTML]{949494}74\% & \cellcolor[HTML]{AEAEAE}61\% & \cellcolor[HTML]{7F7F7F}85\% & \cellcolor[HTML]{767676}89\% \\
 & openISSUE & \cellcolor[HTML]{CFCFCF}44\% & \cellcolor[HTML]{E4E4E4}27\% & \cellcolor[HTML]{E4E4E4}27\% & \cellcolor[HTML]{C3C3C3}51\% & \cellcolor[HTML]{D0D0D0}43\% & \cellcolor[HTML]{C1C1C1}52\% & \cellcolor[HTML]{C2C2C2}51\% & \cellcolor[HTML]{939393}75\% & \cellcolor[HTML]{8F8F8F}77\% & \cellcolor[HTML]{979797}73\% \\
\multirow{-6}{*}{12th} & closedISSUE & \cellcolor[HTML]{D6D6D6}38\% & \cellcolor[HTML]{EBEBEB}21\% & \cellcolor[HTML]{E9E9E9}23\% & \cellcolor[HTML]{D5D5D5}39\% & \cellcolor[HTML]{CFCFCF}44\% & \cellcolor[HTML]{D0D0D0}43\% & \cellcolor[HTML]{6A6A6A}95\% & \cellcolor[HTML]{A5A5A5}66\% & \cellcolor[HTML]{6D6D6D}94\% & \cellcolor[HTML]{787878}88\% \\ \hline
 & \cellcolor[HTML]{FFFFFF}median & \cellcolor[HTML]{FFFFFF}41\% & \cellcolor[HTML]{FFFFFF}27\% & \cellcolor[HTML]{FFFFFF}28\% & \cellcolor[HTML]{FFFFFF}41\% & \cellcolor[HTML]{FFFFFF}44\% & \cellcolor[HTML]{FFFFFF}50\% & \cellcolor[HTML]{FFFFFF}69\% & \cellcolor[HTML]{FFFFFF}60\% & \cellcolor[HTML]{FFFFFF}75\% & \cellcolor[HTML]{FFFFFF}75\%
\end{tabular}}
\end{table}

For SA results, as we see in \tbl{win_sa}, although the median win rate of hyperparameter tuned methods decreased a bit, they still outperform all the untuned methods. Specifically, ASKL wins from $57\%$ to $82\%$, and DECAERT wins from $44\%$ to $84\%$ out of 4 different prediction ways on 1,159 projects. Compare to KNN wins from $21\%$ to $57\%$, LNR wins from $15\%$ to $48\%$, SVR wins from $19\%$ to $41\%$, RFT wins from $27\%$ to $58\%$, CART wins from $29\%$ to $62\%$, and tuned method RDCART wins from $38\%$ to $62\%$, GSCART wins from $45\%$ to $84\%$ and FLASH wins from $39\%$ to $78\%$, respectively. In most cases, the winning rates of the untuned methods  are less than $40\%$. After we take a further look, SVR and LNR performs relatively worse, the median winning rate is only $25\%$ and $26\%$, respectively.

Based on the results from our experiments, we conclude that:

\begin{blockquote}
\noindent
\textbf{Answer 3}: 
Hyperparameter optimized methods generate better prediction performance than the other methods.
\end{blockquote}

As state above, future work
might discover better optimizers
(for health indicator prediction) than our current methods. That said, these {\bf RQ3}
results tell us that HPO can find models that make better predictions  than many other approaches (that are used widely in the   literature).

\begin{table}[!t]
\centering
\caption{Statistical analysis of the SA results: the win rate (ranked first using the Friedman Nemenyi test of Section~\ref{sect:stats} for the    different treatments).
Raw performance measured in terms of SA for predictions $N$ months into the future.}
\label{tbl:win_sa} 
\resizebox{0.97\textwidth}{!}{
\begin{tabular}{llcccccccccc}
\multicolumn{1}{c}{Months} & \multicolumn{1}{c}{Health Indicator} & KNN & LNR & SVR & RFT & CART & RDCART & GSCART & FLASH & DECART & ASKL \\ \hline
 & commit & \cellcolor[HTML]{D0D0D0}44\% & \cellcolor[HTML]{EFEFEF}18\% & \cellcolor[HTML]{E3E3E3}28\% & \cellcolor[HTML]{DBDBDB}34\% & \cellcolor[HTML]{D5D5D5}40\% & \cellcolor[HTML]{CDCDCD}46\% & \cellcolor[HTML]{9B9B9B}71\% & \cellcolor[HTML]{9F9F9F}69\% & \cellcolor[HTML]{8E8E8E}77\% & \cellcolor[HTML]{979797}73\% \\
 & contributor & \cellcolor[HTML]{B6B6B6}57\% & \cellcolor[HTML]{DBDBDB}35\% & \cellcolor[HTML]{E1E1E1}30\% & \cellcolor[HTML]{CECECE}45\% & \cellcolor[HTML]{D6D6D6}38\% & \cellcolor[HTML]{D1D1D1}42\% & \cellcolor[HTML]{AFAFAF}61\% & \cellcolor[HTML]{C5C5C5}50\% & \cellcolor[HTML]{A0A0A0}68\% & \cellcolor[HTML]{8C8C8C}78\% \\
 & openPR & \cellcolor[HTML]{CACACA}47\% & \cellcolor[HTML]{EBEBEB}22\% & \cellcolor[HTML]{E3E3E3}28\% & \cellcolor[HTML]{D0D0D0}43\% & \cellcolor[HTML]{CACACA}47\% & \cellcolor[HTML]{C8C8C8}48\% & \cellcolor[HTML]{9F9F9F}69\% & \cellcolor[HTML]{B2B2B2}59\% & \cellcolor[HTML]{9C9C9C}70\% & \cellcolor[HTML]{9F9F9F}69\% \\
 & closePR & \cellcolor[HTML]{DADADA}35\% & \cellcolor[HTML]{E2E2E2}29\% & \cellcolor[HTML]{D4D4D4}40\% & \cellcolor[HTML]{D1D1D1}42\% & \cellcolor[HTML]{CFCFCF}44\% & \cellcolor[HTML]{CDCDCD}45\% & \cellcolor[HTML]{C0C0C0}52\% & \cellcolor[HTML]{B6B6B6}57\% & \cellcolor[HTML]{C3C3C3}51\% & \cellcolor[HTML]{AFAFAF}61\% \\
 & openISSUE & \cellcolor[HTML]{C9C9C9}48\% & \cellcolor[HTML]{E7E7E7}25\% & \cellcolor[HTML]{EFEFEF}19\% & \cellcolor[HTML]{D3D3D3}41\% & \cellcolor[HTML]{D1D1D1}42\% & \cellcolor[HTML]{D6D6D6}39\% & \cellcolor[HTML]{9A9A9A}71\% & \cellcolor[HTML]{8C8C8C}78\% & \cellcolor[HTML]{8F8F8F}77\% & \cellcolor[HTML]{909090}76\% \\
\multirow{-6}{*}{1st} & closedISSUE & \cellcolor[HTML]{D3D3D3}41\% & \cellcolor[HTML]{EEEEEE}19\% & \cellcolor[HTML]{E8E8E8}24\% & \cellcolor[HTML]{DCDCDC}34\% & \cellcolor[HTML]{D7D7D7}38\% & \cellcolor[HTML]{C4C4C4}50\% & \cellcolor[HTML]{C6C6C6}49\% & \cellcolor[HTML]{D2D2D2}42\% & \cellcolor[HTML]{CACACA}47\% & \cellcolor[HTML]{B7B7B7}57\% \\ \hline
 & commit & \cellcolor[HTML]{E2E2E2}29\% & \cellcolor[HTML]{EAEAEA}22\% & \cellcolor[HTML]{DCDCDC}34\% & \cellcolor[HTML]{BBBBBB}55\% & \cellcolor[HTML]{D1D1D1}43\% & \cellcolor[HTML]{CFCFCF}44\% & \cellcolor[HTML]{BBBBBB}55\% & \cellcolor[HTML]{A1A1A1}68\% & \cellcolor[HTML]{A6A6A6}65\% & \cellcolor[HTML]{9B9B9B}71\% \\
 & contributor & \cellcolor[HTML]{C7C7C7}49\% & \cellcolor[HTML]{E7E7E7}24\% & \cellcolor[HTML]{ECECEC}21\% & \cellcolor[HTML]{DCDCDC}34\% & \cellcolor[HTML]{D2D2D2}41\% & \cellcolor[HTML]{D2D2D2}41\% & \cellcolor[HTML]{9D9D9D}70\% & \cellcolor[HTML]{CFCFCF}44\% & \cellcolor[HTML]{989898}72\% & \cellcolor[HTML]{9B9B9B}71\% \\
 & openPR & \cellcolor[HTML]{EBEBEB}21\% & \cellcolor[HTML]{E1E1E1}30\% & \cellcolor[HTML]{E3E3E3}28\% & \cellcolor[HTML]{DBDBDB}34\% & \cellcolor[HTML]{D7D7D7}37\% & \cellcolor[HTML]{D3D3D3}41\% & \cellcolor[HTML]{CECECE}45\% & \cellcolor[HTML]{B1B1B1}60\% & \cellcolor[HTML]{A9A9A9}64\% & \cellcolor[HTML]{A6A6A6}65\% \\
 & closePR & \cellcolor[HTML]{E2E2E2}28\% & \cellcolor[HTML]{D4D4D4}40\% & \cellcolor[HTML]{D3D3D3}41\% & \cellcolor[HTML]{D9D9D9}36\% & \cellcolor[HTML]{ADADAD}62\% & \cellcolor[HTML]{C3C3C3}51\% & \cellcolor[HTML]{A7A7A7}65\% & \cellcolor[HTML]{A4A4A4}66\% & \cellcolor[HTML]{8C8C8C}78\% & \cellcolor[HTML]{868686}81\% \\
 & openISSUE & \cellcolor[HTML]{D7D7D7}38\% & \cellcolor[HTML]{C9C9C9}48\% & \cellcolor[HTML]{E7E7E7}24\% & \cellcolor[HTML]{D3D3D3}41\% & \cellcolor[HTML]{D0D0D0}43\% & \cellcolor[HTML]{CBCBCB}47\% & \cellcolor[HTML]{A8A8A8}64\% & \cellcolor[HTML]{CDCDCD}45\% & \cellcolor[HTML]{CFCFCF}44\% & \cellcolor[HTML]{B2B2B2}59\% \\
\multirow{-6}{*}{3rd} & closedISSUE & \cellcolor[HTML]{DCDCDC}34\% & \cellcolor[HTML]{E8E8E8}24\% & \cellcolor[HTML]{ECECEC}21\% & \cellcolor[HTML]{D0D0D0}43\% & \cellcolor[HTML]{D2D2D2}41\% & \cellcolor[HTML]{B7B7B7}57\% & \cellcolor[HTML]{959595}74\% & \cellcolor[HTML]{A7A7A7}65\% & \cellcolor[HTML]{868686}81\% & \cellcolor[HTML]{A5A5A5}66\% \\ \hline
 & commit & \cellcolor[HTML]{DBDBDB}34\% & \cellcolor[HTML]{E8E8E8}24\% & \cellcolor[HTML]{E9E9E9}23\% & \cellcolor[HTML]{DEDEDE}32\% & \cellcolor[HTML]{C5C5C5}50\% & \cellcolor[HTML]{D0D0D0}43\% & \cellcolor[HTML]{ABABAB}63\% & \cellcolor[HTML]{BEBEBE}53\% & \cellcolor[HTML]{9D9D9D}70\% & \cellcolor[HTML]{959595}74\% \\
 & contributor & \cellcolor[HTML]{D0D0D0}43\% & \cellcolor[HTML]{DDDDDD}32\% & \cellcolor[HTML]{DDDDDD}33\% & \cellcolor[HTML]{E4E4E4}27\% & \cellcolor[HTML]{CBCBCB}47\% & \cellcolor[HTML]{C8C8C8}48\% & \cellcolor[HTML]{BCBCBC}54\% & \cellcolor[HTML]{CECECE}45\% & \cellcolor[HTML]{ACACAC}62\% & \cellcolor[HTML]{A2A2A2}67\% \\
 & openPR & \cellcolor[HTML]{D1D1D1}42\% & \cellcolor[HTML]{D4D4D4}40\% & \cellcolor[HTML]{E7E7E7}24\% & \cellcolor[HTML]{D6D6D6}38\% & \cellcolor[HTML]{D0D0D0}43\% & \cellcolor[HTML]{D0D0D0}43\% & \cellcolor[HTML]{ABABAB}63\% & \cellcolor[HTML]{C5C5C5}50\% & \cellcolor[HTML]{808080}84\% & \cellcolor[HTML]{8C8C8C}78\% \\
 & closePR & \cellcolor[HTML]{E6E6E6}25\% & \cellcolor[HTML]{E0E0E0}30\% & \cellcolor[HTML]{E5E5E5}27\% & \cellcolor[HTML]{D9D9D9}36\% & \cellcolor[HTML]{D3D3D3}41\% & \cellcolor[HTML]{BABABA}55\% & \cellcolor[HTML]{888888}80\% & \cellcolor[HTML]{A6A6A6}65\% & \cellcolor[HTML]{9A9A9A}71\% & \cellcolor[HTML]{909090}76\% \\
 & openISSUE & \cellcolor[HTML]{D5D5D5}40\% & \cellcolor[HTML]{F0F0F0}17\% & \cellcolor[HTML]{EBEBEB}21\% & \cellcolor[HTML]{DDDDDD}33\% & \cellcolor[HTML]{D0D0D0}43\% & \cellcolor[HTML]{C7C7C7}49\% & \cellcolor[HTML]{C2C2C2}51\% & \cellcolor[HTML]{C3C3C3}51\% & \cellcolor[HTML]{AFAFAF}61\% & \cellcolor[HTML]{959595}74\% \\
\multirow{-6}{*}{6th} & closedISSUE & \cellcolor[HTML]{D6D6D6}38\% & \cellcolor[HTML]{E4E4E4}27\% & \cellcolor[HTML]{E8E8E8}24\% & \cellcolor[HTML]{DDDDDD}33\% & \cellcolor[HTML]{D5D5D5}39\% & \cellcolor[HTML]{D0D0D0}43\% & \cellcolor[HTML]{C2C2C2}51\% & \cellcolor[HTML]{D5D5D5}39\% & \cellcolor[HTML]{C3C3C3}51\% & \cellcolor[HTML]{ABABAB}63\% \\ \hline
 & commit & \cellcolor[HTML]{D6D6D6}39\% & \cellcolor[HTML]{F3F3F3}15\% & \cellcolor[HTML]{ECECEC}21\% & \cellcolor[HTML]{CDCDCD}45\% & \cellcolor[HTML]{E1E1E1}29\% & \cellcolor[HTML]{ADADAD}62\% & \cellcolor[HTML]{AFAFAF}61\% & \cellcolor[HTML]{BFBFBF}53\% & \cellcolor[HTML]{989898}72\% & \cellcolor[HTML]{A0A0A0}68\% \\
 & contributor & \cellcolor[HTML]{B7B7B7}57\% & \cellcolor[HTML]{E9E9E9}23\% & \cellcolor[HTML]{E7E7E7}25\% & \cellcolor[HTML]{DDDDDD}32\% & \cellcolor[HTML]{D6D6D6}39\% & \cellcolor[HTML]{D2D2D2}41\% & \cellcolor[HTML]{C3C3C3}51\% & \cellcolor[HTML]{A6A6A6}65\% & \cellcolor[HTML]{B4B4B4}58\% & \cellcolor[HTML]{A2A2A2}67\% \\
 & openPR & \cellcolor[HTML]{DBDBDB}34\% & \cellcolor[HTML]{E5E5E5}26\% & \cellcolor[HTML]{E4E4E4}27\% & \cellcolor[HTML]{CDCDCD}45\% & \cellcolor[HTML]{C3C3C3}51\% & \cellcolor[HTML]{C2C2C2}51\% & \cellcolor[HTML]{8C8C8C}78\% & \cellcolor[HTML]{989898}72\% & \cellcolor[HTML]{B4B4B4}58\% & \cellcolor[HTML]{929292}75\% \\
 & closePR & \cellcolor[HTML]{CFCFCF}44\% & \cellcolor[HTML]{E9E9E9}23\% & \cellcolor[HTML]{DBDBDB}35\% & \cellcolor[HTML]{D4D4D4}40\% & \cellcolor[HTML]{D7D7D7}37\% & \cellcolor[HTML]{D6D6D6}38\% & \cellcolor[HTML]{B5B5B5}58\% & \cellcolor[HTML]{AEAEAE}61\% & \cellcolor[HTML]{A5A5A5}66\% & \cellcolor[HTML]{9B9B9B}71\% \\
 & openISSUE & \cellcolor[HTML]{D1D1D1}43\% & \cellcolor[HTML]{E3E3E3}28\% & \cellcolor[HTML]{E7E7E7}25\% & \cellcolor[HTML]{B4B4B4}58\% & \cellcolor[HTML]{C9C9C9}48\% & \cellcolor[HTML]{C9C9C9}48\% & \cellcolor[HTML]{818181}84\% & \cellcolor[HTML]{BABABA}55\% & \cellcolor[HTML]{949494}74\% & \cellcolor[HTML]{979797}73\% \\
\multirow{-6}{*}{12th} & closedISSUE & \cellcolor[HTML]{DEDEDE}32\% & \cellcolor[HTML]{E0E0E0}31\% & \cellcolor[HTML]{E9E9E9}23\% & \cellcolor[HTML]{D9D9D9}36\% & \cellcolor[HTML]{DADADA}35\% & \cellcolor[HTML]{CDCDCD}46\% & \cellcolor[HTML]{BABABA}55\% & \cellcolor[HTML]{ADADAD}62\% & \cellcolor[HTML]{818181}84\% & \cellcolor[HTML]{858585}82\% \\ \hline
 & median & 39\% & 26\% & 25\% & 37\% & 42\% & 46\% & 62\% & 58\% & 69\% & 71\%
\end{tabular}
 }
\end{table}

\section{Discussion}
\label{sect:discu}


In this section, we look into the efficiency of our methods, and discuss the potential issues observed from experiment results.
 
\subsection{The efficiency of hyperparameter optimization}
\label{sect:runtime}

\begin{table}[!b]
\centering
\begin{adjustbox}{max width=1\textwidth}
\begin{tabular}{l|c|c|c|c}
\rowcolor[HTML]{BDBDBD} 
 & RDCART & GSCART & FLASH & DECART \\ \hline
\rowcolor[HTML]{FFFFFF} 
min & 6 seconds & 6 seconds & 4 seconds & 7 seconds \\
\rowcolor[HTML]{F3F3F3} 
max & 17 seconds & 19 seconds & 27 seconds & 24 seconds \\
\rowcolor[HTML]{FFFFFF} 
mean & 13 seconds & 12 seconds & 13 seconds & 14 seconds \\
\rowcolor[HTML]{F3F3F3} 
median & 11 seconds & 10 seconds & 11 seconds & 12 seconds 
\end{tabular}
\end{adjustbox}
\caption{  Runtimes of hyperparameter tuned methods for a single project.   }
\label{tbl:runtime}
\end{table}

In the experiments, our hyperparameter tuned methods are not only effective (as shown in \tbl{win_mre} and \tbl{win_sa}), but also  very fast. \tbl{runtime} shows the min, max, mean and median runtime of four hyperparameter tuned methods for a single project.
The mean runtime of each project are similar for these methods (12 to 14 seconds). However, methods like FLASH and DECART 
use more time  (in the median case, 27 and 24 seconds, respectively). This time includes optimizing CART for each 
specific dataset, and then making predictions. Note that, for these experiments, we make no use of any special hardware (i.e. we used neither GPUs nor cloud services that interleave multiple cores in some clever manner).

Also, as previously mentioned, for ASKL, we restrict its runtime for each project to 15 seconds (the upper bound of mean runtime of other hyperparameter optimized predictors). Without this restriction, ASKL can benefit from longer runtime to keep finding better solutions. We randomly select 20 from 1,159 projects and test the performance of ASKL without additional runtime restriction (its default time limit is 3,600 seconds) and DECART, and the result is summarized in \tbl{askl}. In this test, we find that DECART can reach similar prediction performance as ASKL, with much faster runtime (324 seconds vs 1,843 seconds). In a result that is somewhat troubling for ASKL,
that algorithm seems to have occasionally large outlier runs, suggesting
that it struggles sometimes to find solutions. Meanwhile, DECART's
mean and max runtimes are very similar, suggesting that the performance of this algorithm might be more stable across a wider range of problems.


The efficiency of hyperparameter optimization in health indicators prediction is an important finding. 
In our experience, the complexity of hyperparameter optimization is a major concern that limits its widespread use in different domains. For example, in a defect prediction study, Fu et al. report that hyperparameter optimization for code defect prediction can use
up to nearly three days of CPU per dataset~\cite{Fu2016TuningFS}.
If our 1,000+ projects required the similar amount of CPU resources, then it would be a major blocker to the use of the proposed methods in this paper.

But why is methods like DECART runs this fast and effective?
Firstly,  our methods work on  relatively small datasets. 
This paper studies three to five years of project data. For each month, we  extract the 10 features shown in  Table~\ref{tbl:feature}. That is to say, hyperparameter tuned methods only have to explore datasets up to $\mathit{10*60}$ data points per project.

Secondly, as to why are hyperparameter tuned methods so effective, we note that many data mining algorithms  rely on  statistical properties that are emergent in   large samples of data~\cite{witten11}. Hence they  
have problems reasoning about  datasets with only $\mathit{10*60}$ data points.  
Accordingly, to enable effective data mining,
it is important to adjust the learners to the   idiosyncrasies of the dataset 
(via hyperparameter optimization).

\begin{table}[!t]
\centering
\caption{  Runtime and performance comparison between ASKL with   less
restrictive runtimes and DECART, on 20 randomly selected projects. }
\label{tbl:askl}
\begin{adjustbox}{max width=1\textwidth}
\begin{tabular}{c|c|c}
method & DECART & ASKL \\ \hline
median MRE & 0.35 & 0.34 \\
runtime max & 20 seconds & 154 seconds \\
runtime mean & 16 seconds & 92 seconds \\
runtime total & 324 seconds & 1843 seconds
\end{tabular}
\end{adjustbox}
\end{table}

\subsection{On other time predictions}
\label{sect:other}
In our experiment, we observe that when predicting specific health indicators, some HPO methods can achieve 0\% error in some cases. Such zero error is a red flag that needs to be investigated since they might be due to overfitting or programming errors (such as use the test value as both the predicted and actual value for the MRE calculation). What we found was that the older the project, the less the programmer activity.
  Hence,  it is hardly surprising that good learners could correctly predict (e.g.) zero closed pull requests.

But that raises another red flag: suppose {\em all} our projects had reached some steady state prior to April 2020. In that case, predicting (say) the next month's value of health indicator would be a simple matter of repeating last month's value. In our investigation, we have three reasons for believing that this is not the case.
Firstly, prediction in this domain is difficult. If such steady state had been achieved, then all our learners would be reporting very low errors. As seen in Table~\ref{tbl:med_mre}, this is not the case.

Secondly, we looked into the columns in our raw data, looking for long sequences of stable or zero values. This case does not happen in most cases: our data contains many variations across the entire lifecycle of our projects.

Thirdly, just to be sure, we conducted another round of experiments. Instead of predicting for the most recent months, we do the prediction for an earlier period using
data collected prior to that time point.
Table~\ref{tbl:mid} shows the results. In this table, if a project had (say) $N=60$ months of data, we went to months $N/2$ and used DECART to predicted 12 months into the future (to $N/2+12$). The columns for Table~\ref{tbl:mid}  should be compared to the right-hand-side columns of Table~\ref{tbl:med_mre}, Table~\ref{tbl:iqr_mre},  Table~\ref{tbl:med_sa}, and Table~\ref{tbl:iqr_sa}. In that comparison, we see that predicting for months in mid period can
generate comparable results as predicting for most recent months.
 
In summary, our results are not unduly biased by predicting just for the recent months. As the evidence, we can still obtain accurate results if we predict for earlier months.

\begin{table}[!t]
\centering
\caption{The performance of DECART, staring mid-way through a project, then predicting 12 months into the future.}
\label{tbl:mid}
\begin{adjustbox}{max width=0.98\textwidth}
\begin{tabular}{lrrrr}
 & \multicolumn{1}{c}{Median MRE} & \multicolumn{1}{c}{IQR MRE} & \multicolumn{1}{c}{Median SA} & \multicolumn{1}{c}{IQR SA} \\
commit & \cellcolor[HTML]{F3F3F3}70\% & \cellcolor[HTML]{F3F3F3}119\% & \cellcolor[HTML]{F3F3F3}31\% & \cellcolor[HTML]{F3F3F3}98\% \\
contributor & \cellcolor[HTML]{FFFFFF}45\% & \cellcolor[HTML]{FFFFFF}87\% & \cellcolor[HTML]{FFFFFF}38\% & \cellcolor[HTML]{FFFFFF}147\% \\
openPR & \cellcolor[HTML]{F3F3F3}37\% & \cellcolor[HTML]{F3F3F3}85\% & \cellcolor[HTML]{F3F3F3}36\% & \cellcolor[HTML]{F3F3F3}112\% \\
closePR & \cellcolor[HTML]{FFFFFF}66\% & \cellcolor[HTML]{FFFFFF}93\% & \cellcolor[HTML]{FFFFFF}25\% & \cellcolor[HTML]{FFFFFF}93\% \\
openISSUE & \cellcolor[HTML]{F3F3F3}60\% & \cellcolor[HTML]{F3F3F3}67\% & \cellcolor[HTML]{F3F3F3}30\% & \cellcolor[HTML]{F3F3F3}154\% \\
closedISSUE & \cellcolor[HTML]{FFFFFF}34\% & \cellcolor[HTML]{FFFFFF}65\% & \cellcolor[HTML]{FFFFFF}38\% & \cellcolor[HTML]{FFFFFF}111\%
\end{tabular}
\end{adjustbox}
\end{table}

\section{Threats to validity}
\label{sect:threa}
The design of this study may have several validity threats~\cite{feldt2010validity}. The following issues should be considered to avoid jeopardizing conclusions made from this work.

\textbf{Parameter Bias:} The settings to control the hyperparameters of the prediction methods can have a positive effect on the efficacy of the prediction. By using hyperparameter optimized method in our experiment, we explore the space of possible hyperparameters for the predictor, hence we assert that this study suffers less parameter bias than some other studies.

\textbf{Survey Bias:} To verify whether our potential health indicators actually matter, we made a survey to open source project developers to ask about their opinions based on their development experience. While most features were considered to be relevant to the project health in the survey, we would not claim the result was a complete set of project health indicators. In our survey, the choice of features was limited to several options in order to keep it easy to respond. Although the participants could provide additional thoughts in the following question, this would still narrow down their opinions. In future, additional knowledge from participants will be added to reduce this bias. 

\textbf{Metric Bias:} We use Magnitude of the Relative Error (MRE) as one of the performance metrics in the experiment. However, MRE is criticized because of its bias towards error underestimations~\cite{foss2003simulation,kitchenham2001accuracy,korte2008confidence,port2008comparative,shepperd2000building,stensrud2003further}. Specifically, when the benchmark error is small or equal to zero, the relative error could become extremely large or infinite. This may lead to an undefined mean or at least a distortion of the result~\cite{chen2017new}. In our study, we do not abandon MRE since there exist known baselines for human performance in effort estimation expressed in terms of MRE~\cite{Jorgensen03}. To overcome this limitation, we set a customized MRE treatment to deal with ``divide by zero'' issue and also apply Standardized Accuracy (SA) as the other measure of the performance.

\textbf{Sampling Bias:} 
In our study, we collect 64,181 months with 10 features of 1,159 GitHub projects data for the experiment, and use 6 GitHub development features as health indicators of open-source project. While we reach good prediction performance on those data, it would be inappropriate to conclude that our technique always gets positive result on open-source projects, or the health indicators we use could completely decide the project's health status. 
Another confounding factor is, since the projects we collected have different sizes, domains, life-cycles, etc., they could have different factors regarding the predicting performance of health indicators.
Also, in the study, we focus on the active project for the data collection. Those excluded inactive repositories might also provide useful data about how projects failed then give more exemplars for model training.
To mitigate these problems, we release open source resources of our work to support the research community to reproduce, improve or refute our results on broader data and indicators.

\section{Conclusion and Future Work}
\label{sect:concl}

Our results make a compelling case for open source software projects. 
Software developed on some public platforms is a source of data that can be used to make accurate predictions  about those projects. While the activity of a single developer may be random and hard to predict, when large groups of developers work together on software projects,
the resulting behavior can be predicted with good accuracy. For example, after building predictors for six project health indicators, we can make predictions with low error rates (median values usually under 25\%). 

Our results come with some caveats. The patterns of some activities are harder to be learned, for the law of large numbers. We know this since we cannot constantly get high accuracy in predictions. For example, 
across our six health indicators, the predicting performances of closePR and commit are not as good as when predicting the number of contributors (as shown in Table~\ref{tbl:med_mre}). 
Also, to make predictions, we must take care to tune the data mining algorithms to the idiosyncrasies of the datasets.
Some data mining algorithms rely on statistical properties that are emergent in large samples of data.
Hence, such algorithms may have problems reasoning about very small datasets, such as those studied here.
Hence, before making predictions, it is vitally   important to adjust the learners to the idiosyncrasies of the dataset via hyperparameter optimization. Unlike prior hyperparameter optimization work by Fu et al.~\cite{Fu2016TuningFS}, our optimization process is very fast (a few seconds per dataset). Accordingly, we assert that for predicting project health indicators, hyperparameter optimization is the preferred technology.
 
As to future work, there is still much to do. 
Firstly, we know many organizations such as IBM that run large in-house ecosystems where, behind firewalls, thousands of programmers build software using  a private GitHub system. It would be insightful to see if our techniques work for such ``private'' GitHub networks.
Secondly, our results still have large space to improve. Some prediction tasks are harder than others
(e.g. commits, closed PR).
In our study, DE has shown good prediction performance comparing to other methods, exploring other evolutionary algorithms~\cite{wu2018ensemble,das2016recent} on different learners (e.g. random forest) or applying auto-sklearn~\cite{feurer2019auto} could be useful and might bring even better results.

Further, as to more indicators, there are more practices from real-world business-level cases to explore. 
In our survey, some of the participants mentioned other features they think are relevant to open source project health, we will assess their opinions to find more indicators. 
Also, it would be worth trying if we can derive more effective sophisticated health indicators, such as:

\bi
\item
The number of new joining/leaving contributors.
\item
The change in number of developing features (commits, contributors, openPR, etc.) over time.
\ei

Lastly, an enriched data collection with more features from more types of repositories (e.g. inactive or archived projects) would be helpful for our model learning. With enough training data, we could also explore deep neural network methods (e.g. LSTM) on projects from thousands of repositories, and try to detect anomalies in their developments that may jeopardize the health of these software projects.

\section*{Acknowledgements}
This work is partially funded by a National Science Foundation
Grant \#1703487.

\bibliographystyle{ACM-Reference-Format}
\bibliography{bibreference}

\end{document}